\documentclass[lettersize,journal]{IEEEtran}
\usepackage{amsmath,amsfonts}
\usepackage{algorithmic}
\usepackage{array}
\usepackage[caption=false,font=normalsize,labelfont=sf,textfont=sf]{subfig}
\usepackage{textcomp}
\usepackage{stfloats}
\usepackage{url}
\usepackage{verbatim}
\usepackage{graphicx}
\def\BibTeX{{\rm B\kern-.05em{\sc i\kern-.025em b}\kern-.08em
    T\kern-.1667em\lower.7ex\hbox{E}\kern-.125emX}}
\usepackage{balance}

\usepackage{cite}

\usepackage{amsmath}
\usepackage{xcolor}
\usepackage[utf8]{inputenc}
\usepackage{hyperref}
\usepackage{svg}

\newcommand{\orcidlink}[1]{\href{https://orcid.org/#1}{\includegraphics[width=10pt]{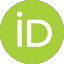}}}

\definecolor{mustardyellow}{RGB}{210, 178, 2} 

\definecolor{forestgreen}{rgb}{0.0, 0.5, 0.0}

\usepackage{booktabs}
\usepackage{lipsum}
\usepackage{multirow}
\usepackage{threeparttable}
\usepackage{xcolor,colortbl}
\newcommand{\gray}[0]{\cellcolor{gray!50}}
\usepackage{svg}

\usepackage{listings}
\lstdefinelanguage{code}{
    basicstyle=\normalfont\ttfamily,
    numberstyle=\scriptsize,
    numbers=left,
    stepnumber=1,
    numbersep=0pt,
    showstringspaces=false,
    breaklines=true,
    frame=lines,
}

\usepackage{amsmath,amsfonts,amssymb}

\begin{document}

\title{Quantum-Safe integration of TLS in SDN networks}

\author{
    Jaime S. Buruaga\IEEEauthorrefmark{1},
    Ruben B. M\'endez\IEEEauthorrefmark{1}, 
    Juan P. Brito\IEEEauthorrefmark{2}, 
    Vicente Martin\IEEEauthorrefmark{2}\\
    \IEEEauthorblockA{\IEEEauthorrefmark{1}Center for Computational Simulation, Universidad Polit\'ecnica de Madrid, Madrid, Spain}\\
    \IEEEauthorblockA{\IEEEauthorrefmark{2}Center for Computational Simulation and DLSIIS, ETSI Inform\'aticos, Universidad Polit\'ecnica de Madrid, Madrid, Spain}\\
    \IEEEcompsocitemizethanks{\IEEEcompsocthanksitem Jaime Sáez de Buruaga is with the Center for Computational Simulation, Universidad Polit\'ecnica de Madrid, Spain, 28223.
    \protect\\
    E-mail: j.saezdeburuaga@upm.es}
}


\maketitle

\begin{abstract}
Shor's algorithm efficiently solves factoring and discrete logarithm problems using quantum computers, compromising all public key schemes used today. These schemes rely on assumptions on their computational complexity, which quantum computers can easily bypass. The solutions have to come from new algorithms -- called Post-Quantum Cryptography (PQC) -- or from new methods, such as Quantum Key Distribution (QKD). The former replicate the computational security ideas of classical public key algorithms, while the latter recurs to use the quantum properties of nature, which also brings a mathematical security proof, potentially offering Information-Theoretic Security. To secure data in the future, we must adopt these paradigms. With the speed of quantum computing advancements, the transition to quantum-safe cryptography within the next decade is critical. Delays could expose long-lived confidential data, as current encryption may be broken before its value expires. However, the shift must balance the adoption of new technologies with maintaining proven systems to protect against present and future threats. In this work, we have selected Transport Layer Security, one of the most widely used protocols, as the foundation to hybridize classical, quantum, and post-quantum cryptography in a way suitable for broad adoption in Software-Defined Networking, the most flexible networking paradigm that has been used to deploy integrated quantum-classical networks. To this end, we use standards for QKD key extraction and SDN integration. The purposed implementation is based on the latest version of TLS and demonstrates advanced capabilities such as rekeying and key transport across a large QKD network, while supporting crypto-agility and maintaining backward compatibility through the use of ciphersuites. The performance of this approach has been demonstrated using a deployed production infrastructure.
\end{abstract}


\section{Introduction}\label{sec:Introduction}

Cryptography is nowadays in a fast and deep process of change. The main reason is the discovery of quantum computing and Shor's algorithm, which is capable of factoring numbers and solving the discrete logarithm problem, the two main staples on which current public key cryptography is based. These two problems are used as one-way functions: those that are easy to go in one direction (e.g., the multiplication of two large integers) and very difficult to invert (e.g.,  factor a large integer in its prime factors). The computational complexity of inverting the problem is assumed as the guardian of the secret key in a public cryptosystem \cite{DH}. 

These mathematical problems have allowed us to agree on a secret key between two partners in a communication that share nothing (except the mutual knowledge of their identities), send authenticated messages, or sign electronic documents. The security of our very convenient Public Key Infrastructure (PKI) relies solely in the assumption of the difficulty of solving these problems. 

This assumption was shattered by the discovery of Shor's algorithm in 1994 \cite{shor, SHOR_1994}. However, it is still required the building of a cryptographycally relevant quantum computer (i.e., a quantum computer able to work on a few thousand effective qubits with negligible noise), and soon it was understood that building such a computer is a very difficult task, and PKI continued to be used ignoring this threat. 

Since then, large investments in quantum technologies have been speeding up the process. Building a quantum computer has started to look like an engineering problem that might be achieved in a relatively short time instead of a major physics endeavor with uncharted problems ahead. 

A quick analysis reveals a significant concern: to protect the secrecy of long-lived information -- like governmental data or medical records, that must be kept secret during the life of the person -- it is imperative to start encrypting the information right now, to mitigate the risk that a future quantum computer could decrypt the data while it remains sensitive (``harvest now, decrypt later'').

For this reason, the US National Institute of Standards and Technology (NIST) published a report \cite {NIST_RepOnPQC} and made a call in 2016 \cite{NIST_CFP_2} to the cryptographic community to find a substitute of the current public key cryptosystems, i.e.: to substitute RSA, Diffie-Hellman Key Exchange (DHKE) and Elliptic-curve cryptography (ECC) algorithms by others that could resist the attack of a quantum computer. This opened a complex standardization effort that is still on going \cite{PQC_Calls}.

The proposed algorithms were based on a variety of mathematical problems -- some of them already known since many years, but were not previously used mainly for practical purposes such as slow performance or the production of large keys or cyphertexts -- like lattices, codes, hashes, etc. Out of the original 69 candidates presented to the first round in January 2017 \cite{PQC_Round_1} and the following calls, four were selected in the fourth and final round in 2022 \cite{NIST_8413}: one Key Encapsulation Mechanism (CRYSTALS-Kyber) and three Digital Signature Schemes (CRYSTALS-Dilithium, SPHINCS+ and FALCON) to proceed to standardization, but only three were finally selected as standards on summer 2024. FIPS 203 defines ML-KEM (Module-Lattice-Based Key-Encapsulation-Mechanism) \cite{FIPS203}, which is based on CRYSTALS-Kyber. FIPS 204 defines ML-DSA (Module-Lattice-Based Digital Signature Algorithm) \cite{FIPS204}, which is based on CRYSTALS-Dilithium. Finally, FIPS 205 defines SLH-DSA (Stateless-Hash-Based Digital Signature Algorithm) \cite{FIPS205}, which is based on SPHINCS+. A further call for additional digital signature schemes is still ongoing to increase the variety of problems with the objective of having a substitute if one family of algorithms turns out to be weaker than expected.

These algorithms are known as post-quantum algorithms and are shown to be resistant to Shor's algorithm. However, they have not been proven to be resistant to any possible attack coming from either quantum or classical computers. We are, therefore, once again relaying on assumptions-based computational complexity security, similar to what we had with factoring or discrete-log problems, without hard mathematical security. Maybe with the added caveat that quantum computing is in its infancy, so we are not sure how powerful new quantum computing attacks might be. Note also that classical attacks are also possible, since none of these methods have a security proof, and have not been tested in practice to the same level as the current crypto-algorithms. This uncertainty is one of the main concerns of many researchers in the field. Thus, the strategy of having KEMs and DSS based on different families of mathematical problems, thereby enabling crypto-agility, is a prudent approach from a security point of view. 

It is worth noting that quantum computing is not considered a threat for symmetric cryptography, which is not based on problems with a closed mathematical form, but in processes based on diffusion, which distribute the information in the plaintext randomly in the cyphertext, and confusion, which encodes the original information into some other form in a random way, obscuring the relationship between the secret key and the cyphertext. The goal in the design of symmetric cryptography is that only brute-force attacks apply, and, in this sense, the maximum gain possible with a quantum computer is a square root in the size of the search space using Grover's algorithm \cite{grover} to perform an exhaustive search, a problem easily solved by doubling the size of the symmetric key.

At the same time than providing new tools for cryptanalysis, quantum technologies also provide new capabilities that can enhance cryptographic systems. The best known method is Quantum Key Distribution (QKD) \cite{Gisin_RMP_02,Martin_17}. QKD is essentially a key distribution method that could be regarded as a substitute for a DHKE, where the functionality of the discrete log is performed by a physical process based on the properties of quantum physics. The advantages of using QKD is that its security has been demonstrated: since it is not based on mathematical properties, it is immune to computational attacks, meaning that no assumptions on the computational power of an attacker must be made. Attacks are only feasible through side-channels of the implementation, and these are only possible while the transmission is taking place. This implies that the long-term security profile of QKD is better known than their computationally based counterparts, where advances in computer technology, new cryptanalysis attacks based on new algorithms, mathematical knowledge or new computational models can potentially appear and weaken or even destroy their security.

These new cryptographic primitives must also be seamlessly integrated into existing telecommunications infrastructures, combining them with current security protocols and well-established methods. One of the most widely used security protocols in this context is Transport Layer Security (TLS) \cite{RFC_2246}, which serves as the ``\textit{de facto} standard'' for securing communications in services such as the World Wide Web, email, and modern videoconference systems. Initially released as Secure Sockets Layer (SSL) \cite{SSL_1995} in 1995, it has been continuously revised and improved for nearly three decades. It was renamed to TLS in 1999 when the Internet Engineering Task Force (IETF) focused their efforts on its evolution to become a standard for secure communication over the Internet. Nowadays, almost every secure transaction, query, or request on the Internet relies on TLS for protection.

Software-Defined Networking \cite{SDN_Arch_ONF_o, SDN_arch_ONF, Kreutz_SDN_Survey} is a network paradigm designed with scalability and flexibility in mind, enabling networks to adapt to new devices and evolving needs.

The SDN principles are based on decoupling the network's control and data planes, using standardized open interfaces and a logically centralized SDN controller that has a global view of the network and can control the programmable devices in the data forwarding plane.

By using the SDN paradigm, enterprises and telecommunications companies gain unprecedented programmability, enabling scalable deployment, control, and management through automation. This makes SDN an ideal platform for deploying specialized systems and facilitating the transition to emerging technologies like QKD \cite{Engineering_SDN_QKD, MadQCI}. Since its invention, SDN has been adopted by most major telecommunications companies and has served as a vehicle for the widespread adoption of new technologies.

Our goal here is to combine state of the art conventional and quantum-resistant cryptography, integrating quantum and/or post-quantum primitives with TLS -- as a key representative of secure transport protocols -- and SDN, a major paradigm that also facilitates the deployment of QKD. This approach is done a) using as many standards as possible, in particular on the QKD side, either for key extraction and for SDN integration; b) supporting crypto-agility as well as TLS backwards compatibility through the use of {\sl ciphersuites} c) supporting advanced key provisioning systems for enhanced performance and scalability. This pragmatic approach has been thoroughly tested in real production networks \cite{MadQCI}. The implementation proposed in this work demonstrates how to enable vendor-independent end-to-end security among network nodes using multi-hop QKD key transport and advanced features like rekeying. We argue that only by demonstrating a complete, standards-based solution like the one presented here -- adhering to the principles of SDN, crypto-agility, compatibility and scalability -- we can enable the broad adoption of hybrid infrastructures that integrate conventional, quantum, and post-quantum technologies in quantum-resistant infrastructures. 

The paper is organized as follows; first, we deal with the details of the different technologies in the Background section (\ref{sec:Background}), then we discuss our quantum-safe TLS approach, including the use of standards. The crypto-agile and backward compatible SDN-TLS architecture and its implementation within a real-world are presented next. Finally, the performance results are presented, summarizing the main points in the conclusion and future work. 
\section{Background and state of the art}\label{sec:Background}

\subsection{QKD and PQC}

A QKD process requires the ability to produce, transmit, and measure qubits, the quantum counterparts of the bits. Central to this process are three fundamental quantum principles. The superposition principle asserts that single quantum states can be represented as a linear combination of the set of quantum states that span the space where the qubits are described (e.g., a 2-dimensional Hilbert space for a single qubit). From this principle, the no-cloning theorem follows \cite{no_cloning}, which states that an unknown quantum state cannot be perfectly copied. Additionally, the measurement principle establishes that we make measures using a specific basis of the state space and that, from the underlying linear combination representing the state being measured, 
only one of the states in the linear combination will be observed. The probability of observing each possible outcome is determined by the coefficients of the linear combination, introducing inherent unpredictability.

These physical principles can be leveraged in a protocol that generates a key whose information leakage (i.e., the amount of secret key known outside of the legitimate parties executing the protocol) can be bounded as tightly as desired. The users of the protocol can specify the desired secrecy level, thereby controlling information leakage. The correct execution of the protocol ensures that this level is achieved, regardless of the attacker's resources or computational power \cite{Shor_00, RMP_09}. The worst-case scenario would be the failure to produce a key, the equivalent of a denial-of-service attack. This guarantees security that is independent of the computational power of the attacker, known as Information-Theoretic Security (ITS).

QKD protocols come in different flavors, with one of the main distinctions being how they generate and transmit quantum states. These protocols are categorized as either prepare and measure (PM) or entangled pair (EP) protocols. In PM protocols, quantum states are prepared at the emitter and sent to the receiver, where they are measured. The most well-known PM protocol is BB84 \cite{Bennet_84}. In EP protocols, entangled quantum pairs -- quantum states that hold special correlations characteristics of the quantum realm -- are prepared, typically by a source located somewhere along the communications channel, and then distributed to the two users, with each user receiving one member of the pair. The EP91 \cite{Ekert_91} protocol is the most well-known EP protocol. In this case, it can be demonstrated that the security of the system is independent of the operator of the source: even if the operator is a potential Eavesdropper, they cannot gain any information about the final key established by the users. However, obviously, they can disrupt communication by cutting or blocking the channel. 

The other main characteristic of QKD protocols refers to the type of quantum signal used. The first and most well-known approach uses discrete variables (DV), like when encoding a qubit in, for example, the polarization degree of freedom of a photon. In this case, it is crucial to prepare single-photon states, which means that cooled single-photon detectors have to be used on the other side of the channel.

The other choice is known as continuous variables (CV) \cite{Grosshans_02} and uses the continuous quadratures of the electromagnetic field to carry the quantum signal. In this case more than one photon can potentially be used to encode the single qubit and the detection system is notably different. It does not require single photon detectors, but standard, albeit very low noise, photon detectors like the ones used typically in homodyne or heterodyne detection schemes in optical telecommunications systems. There is no clear winner; DV is more costly because of the detectors, and CV is potentially easier to manufacture, but the post-processing of the signals is much more expensive. 

A third type of QKD system worth mentioning is the device-independent (DI) QKD. These systems use entanglement to reduce the trust required in the hardware, allowing security bounds to be achieved even when devices are not characterized \cite{Prim_23}. The rough idea is to verify the behavior of the QKD system, not by examining its individual components like emitters or detectors, but by subjecting it to a Bell test. Bell tests assess whether a system behaves fully classically or it still holds quantum correlations. If quantum correlations are detected, the system can be used to perform QKD securely; otherwise, it cannot. Technically, achieving full QKD in DI is complex, and currently there is no practical system available. However, a simplified version called Measurement-Device-Independent (MDI) QKD is feasible and can be implemented in telecommunication networks \cite{MDI_12}. MDI-QKD focuses on reducing trust in the measurement devices (detectors) rather than in the entire system. This approach holds significant promise, as it can tolerate higher losses than the standard PM protocols. MDI-QKD  requires the presence of a measurement station, typically located between the emitter and the receiver, which notably does not need to be trusted.

From a network security perspective, all these systems are essentially equivalent, as they ultimately operate as a key agreement protocol between emitter and receiver. From a management point of view, MDI and EP systems usually have an additional component that holds either the detection system (MDI case) or the photon source (EP case) in addition to either two emitter stations in the former case or two detection stations in the latter. Essentially, the intermediate stations do not need to be trusted, as the final shared key is only known by the end stations of the quantum channel, i.e.: both emitters for MDI-type systems or both receivers in the EP case, replicating the situation between emitter and receiver in standard prepare and measure protocols. 

Obviously, the QKD process has its limitations, which are also imposed by the laws of physics. One of the most significant constraints is imposed by absorptions, which limit the maximum distance over which a secure connection can be established \cite{Long-RangeQKD}. Current advances in QKD protocols have extended this distance to up to 600 km, although the resulting key rates are relatively slow. Surpassing this distance requires either a satellite connection \cite{satelites_chinos}, a whole new infrastructure beyond the existing optical fiber one \cite{TuboVacio}, or the deployment of quantum repeaters \cite{Briegel_98}, a technology that is still under development.

Post-Quantum Cryptography (PQC), as mentioned in the Introduction, seeks to replace current public-key algorithms with alternatives that are resistant to quantum computers. Given our limited understanding of quantum computing, this currently translates to resistance against Shor's algorithm. The scope of PQC is thus asymmetric cryptography, in contrast to QKD, which is symmetric. PQC relies on mathematical algorithms executed on standard computing and communications devices, with security grounded in the assumed complexity of a certain mathematical problem, while QKD requires physical channels and devices capable of transmitting, manipulating and detecting quantum signals, with security guaranteed by the fundamental laws of physics. The former can be compromised by computational attacks, while the latter is immune to them, as it is an ITS primitive. However, in both cases, implementations can be vulnerable to side-channel attacks, which exploit flaws in their implementation.

Although PQC algorithms have just recently come to the fore, they have a relatively long history. The PQCrypto conference series started in 2006, and the Quantum-Safe Cryptography workshops, promoted by the European Telecommunications Standards Institute (ETSI), started in 2013. From a mathematical point of view, some of the foundational problems underlying the protocols proposed by NIST for standardization are several decades old and have been extensively discussed in well-established conference series such as CRYPTO, EUROCRYPT or ASIACRYPT. A comprehensive discussion of the five families of problems currently considered as the basis for PQC algorithms includes: lattice-based primitives (finding the shortest/closer vector in a lattice), multivariate-based schemes (solving a system of multivariate polynomial equations), code-based approaches (the decoding problem in a linear code), hash-based functions (finding collisions in cryptographic hash functions), and isogeny-based schemes (finding an isogeny between a pair of supersingular elliptic curves). The classification, along with an assessment framework and a discussion of key sizes for authentication and key establishment, can be found in Ref. \cite{PQC-families}. The algorithms recently standardized by NIST belong to either the lattice family (key encapsulation \cite{FIPS203} and signatures \cite{FIPS204}) or the hash-based family (signatures \cite{FIPS205}). Algorithms based on the other families are also being considered for standardization to comply with the crypto-agility requirements.

However, historically, these algorithms did not compete with current factoring or discrete log-based protocols due to issues like lower computational efficiency, larger key sizes, etc. As a result, they never underwent the rigorous testing in production environments that current protocols have endured, which is a key factor in the trust placed in the security of existing cryptographic schemes. This underscores the importance of the NIST standardization process cited in the Introduction. Its goal is to establish the same confidence level for quantum-safe algorithms that has been achieved through years of practical use of factoring or discrete-log based algorithms. This process represents the first solid step in a transition that will take years to complete. An example of the importance of extensive testing is the case of the isogeny-based algorithm SIKE, which was broken using a portable computer when it had passed to the third round of the NIST process and was close to being standardized. This incident highlights the essential role that thorough testing from both the cryptography research community and end users plays in ensuring the robustness of these new algorithms.

Finally, from a cryptographic perspective, it is essential to consider the evolution from the current quantum-threatened infrastructure to a quantum-safe one, potentially involving a combination of post-quantum algorithms, quantum cryptography, and even existing classical algorithms. In this sense, most security agencies and industrial consortiums recommend a hybrid approach \cite{QED-C_financial}. Hybridization involves using current algorithms that are not quantum-safe in combination with quantum-safe methods, including QKD, for key encapsulation or digital signatures. The final key is derived using a Key Derivation Function (KDF) that combines keys generated by both types of methods. This approach is particularly convenient because the new algorithms still need to demonstrate their resilience against a wide range of attacks -- both cryptanalytical and side-channel attacks -- that are expected to emerge over the coming years. In the absence of security demonstrations, which are only available for QKD and One Time Pad, the trust in the security of these new algorithms is built through extensive testing and analysis. This is also the reason underlying the NIST competition's objective of standardizing algorithms based on qualitatively different mathematical problem families, and what is offered by QKD being a physically, rather than mathematically, based protocol is also qualitatively different. It is also worth noting that these agencies also recommend not relying solely on QKD.

\subsection{SDN}

Software-Defined Networking is a network paradigm designed to address the growing complexity of traditional IP networks. As networks became increasingly difficult to manage, they also became overly dependent on vertically integrated proprietary solutions, which ultimately constrained their evolution and large-scale deployment. SDN seeks to overcome these limitations by enabling greater flexibility in the integration of new devices, automating network management and control, and fostering vendor independence. This approach facilitates the provision of new services on a scale and at a reasonable cost.

The SDN principles are simple (see Ref. \cite{Kreutz_SDN_Survey} for a survey) and can be summarized in the decoupling of control and data planes: In traditional networking, both the control and the data planes are tightly coupled to networking devices (such as switches and routers). 

SDN separates these planes, relying on a logically centralized SDN controller, which acts as the ``brain'' of the network. This controller communicates with programmable network devices using open interfaces: Southbound Application Programming Interfaces (API), instructing them on how to manage traffic based on high-level policies and application requirements, and providing services to network applications via the Northbound API, abstracting the underlying network resources. Several SDN controllers have become prominent in modern telecommunications infrastructures, including open-source solutions like ONOS \cite{ONOS}, OpenDaylight \cite{ODL} or TeraFlow \cite{TeraFlow}, as well as proprietary solutions offered by major networking companies such as Cisco, Nokia, Ciena, etc.
Intensive use of APIs and standards: SDN environments typically utilize software components with specific responsibilities embedded in the network using standards \cite{Schaller_17} and common tools to ease integration. For example, the use of YANG \cite{RFC_7950} models and high levels of abstraction in the management layers are typical. The interfaces used by the logically SDN centralized controller are:

Southbound interfaces: These connect the SDN controller to the network devices. OpenFlow \cite{OpenFlow} was the first protocol to follow this trend and is usually considered the ``father of SDN'', but recently new and modern protocols give more reliability, such as NETCONF \cite{RFC_6241}, RESTCONF \cite{RFC_8040} and gNMI/gRPC \cite{RFC_9232}. These protocols allow the SDN controller to communicate with network elements, like routers and switches, to configure forwarding rules to provision light paths dynamically or set other transport rules.

Northbound interfaces: Enable communication between the SDN controller and network applications, services, or orchestration systems. This allows applications to request specific network behaviors and services without the need of understanding the underlying network infrastructure.

East-West interfaces: The centralized SDN controller can interact with other specialized controllers on the network. Typically, the control plane is shaped by several controllers that cooperate between them, for example the Optical controller, Data Center controllers, WAN controllers, etc. These controllers use East-West APIs to organize and provision different resources that need to be orchestrated. For example, provisioning of the optical path through the optical controllers using transport API for data transmission. 

SDN offers several key advantages over traditional network approaches. The core concept of Centralized Management provides a unified, global view and control of the entire network, streamlining the process of management, configuration, and optimization. The high level of programmability inherent to SDN enables administrators to define and deploy policies dynamically, based on real-time application requirements and traffic patterns, thus simplifying the Operation Systems Support. This programmability also increases flexibility, allowing for the seamless adjustment of network behavior, the integration or removal of network services or hardware, and the modification of policies as needed. In addition, SDN significantly improves network agility, allowing for rapid responses to evolving requirements, traffic changes, and service demands. When combined with Network Function Virtualization, SDN allows rapid deployment of new services, adaptation to emerging technologies, and the ability to respond to security threats or scale resources on demand, which is essential for operators of large-scale infrastructures.

\subsection{TLS and TLS 1.3}\label{subsec:TLS-background}

As mentioned in the Introduction, Transport Layer Security (TLS) \cite{RFC_2246} is widely regarded as the ``\textit{de facto} standard'' for securing communications across a wide range of services.

TLS is implemented on top of a transport layer protocol such as the Transmission Control Protocol and ensures the authenticity, confidentiality, and integrity of the parties involved in the communication through a process known as the TLS handshake. Authentication relies on digital certificates to establish the identity of the server and, optionally, the client that initiated the communication. These certificates are issued by Certificate Authorities (CAs), which serve as trusted entities in the authentication verification process, operating within a PKI. Confidentiality is performed by encrypting the transmitted data, ensuring that it remains unreadable to unauthorized entities. To achieve this, TLS uses a combination of symmetric and asymmetric encryption algorithms to secure authentication and confidentiality. Finally, integrity is guaranteed through cryptographic hash functions and Message Authentication Codes (MACs), ensuring that data remain unaltered during transmission.

TLS has historically been the target of various attacks (e.g., BEAST, POODLE, Triple Handshake, Heartbleed, etc.). To mitigate these threats and improve both security and performance, TLS 1.3 was released in August 2018 under RFC 8446 \cite{RFC_8446}, significantly reducing the number of ciphersuites (from 37 to 5).

Older and less secure cryptographic algorithms and ciphers, such as RSA key transport, finite-field DHKE, SHA-1, or RC4, were removed, and only modern and secure algorithms are supported, such as Authenticated Encryption with Associated Data (AEAD) schemes for symmetric encryption, enhancing overall security. It includes support for perfect forward secrecy (PFS), ensuring that even if a long-term key is compromised, past communications remain secure. Moreover, TLS 1.3 includes mechanisms to mitigate certain types of attacks, such as ``downgrade attack'', where an attacker forces a connection to use an older and less secure version of the protocol. Additionally, TLS 1.3 decreased the number of round-trips required during the handshake process, thereby reducing latency.

The enhancements in TLS 1.3 come at the cost of reduced backward compatibility with the previous version of the protocols, although it includes mechanisms to allow interoperability during the transition period.

Ciphersuites are combinations of cryptographic algorithms designed to secure network connections through TLS. Each ciphersuite defines a specific set of algorithms for key exchange, bulk encryption, and MAC. TLS 1.3 simplifies and strengthens this approach by supporting a reduced set of modern, more secure ciphersuites, such as those based on the Elliptic-curve Diffie-Hellman (ECDH) key exchange, which provides enhanced security. TLS 1.3 also improves session resumption mechanisms, enabling faster reestablishment of secure connections while maintaining robust security guarantees.

In summary, TLS 1.3 is designed to offer a more secure, efficient, and streamlined method for establishing secure connections, addressing the limitations and vulnerabilities found in previous versions of the protocol. Given its widespread use as a core protocol for securing network communications, it is crucial to make it quantum-safe. However, this transition must be approached systematically, preserving the overall design of TLS rather than resorting to quick ad-hoc modifications. Moreover, this integration should be aligned with the evolving network infrastructure requirements of telco operators, who are shifting towards modern network paradigms like SDN. By adopting this approach, it becomes possible to create more flexible and adaptative networks that meet the dynamic demands of modern applications and services, thereby facilitating the broad adoption of a disruptive Quantum-Safe TLS.

\subsubsection{TLS handshake}

Any TLS client-server connection begins with a TLS handshake protocol, which allows both endpoints to securely agree on various parameters used to verify each other's identity and negotiate cryptographic modes, security parameters, and key material for encrypting subsequent communication. TLS has undergone significant changes in the latest version, TLS 1.3. However, since TLS 1.2 (defined in RFC 5246 \cite{RFC_5246}) remains widely used (in May 2024, 29.9\% of the websites surveyed still use TLS 1.2 as the best protocol support, according to Ref. \cite{TLS_usage}), it is important to discuss both versions and highlight the significant differences between them.

The TLS handshake performs two essential functions in secure communication: authentication and key exchange. Authentication ensures the identities of the parties involved, while key exchange facilitates the secure establishment of a symmetric encryption key. Traditionally, authentication has relied on Public Key Cryptography (PKC), which, although effective against classical attacks, are susceptible to quantum attacks. As quantum computing capabilities advance, these vulnerabilities require the adoption of alternative cryptographic methods.

\begin{figure}[htbp]
    \centering
    \includegraphics[width=.97\linewidth]{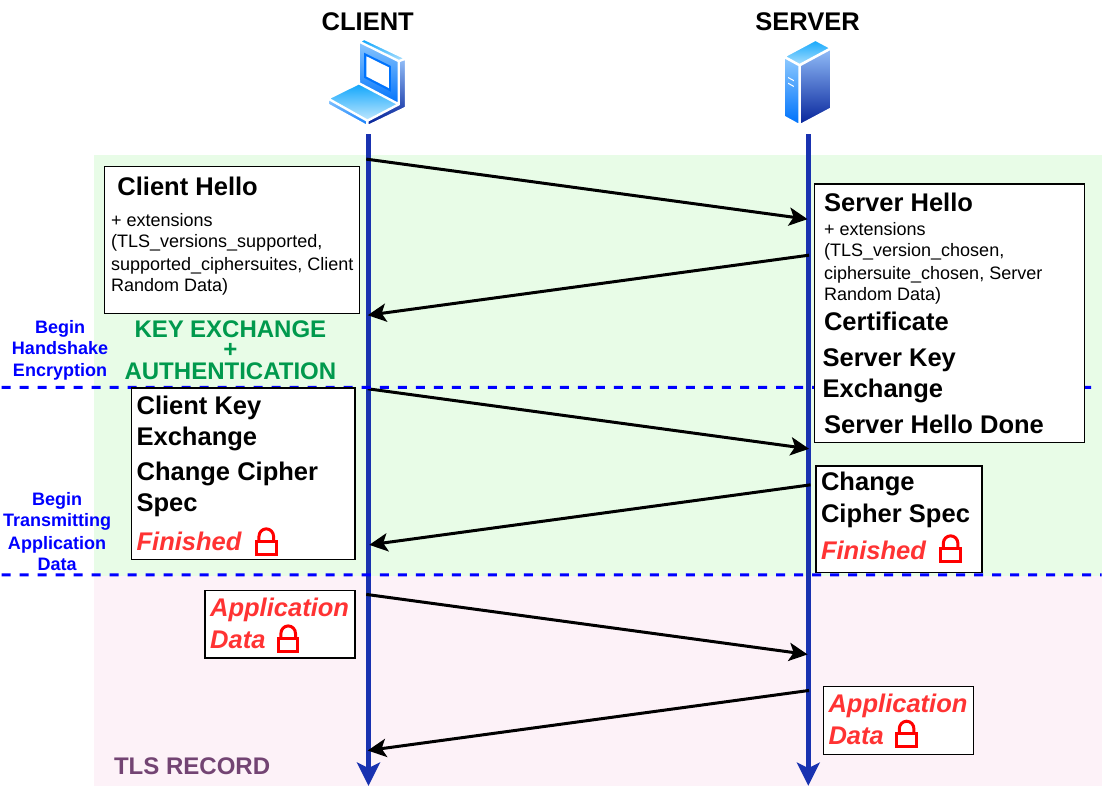}
    \caption{TLS 1.2 handshake protocol (plain messages in black, ciphered messages in red).}
    \label{fig:TLS1.2hs}
\end{figure}

\begin{figure}[htbp]
    \centering
    \includegraphics[width=.97\linewidth]{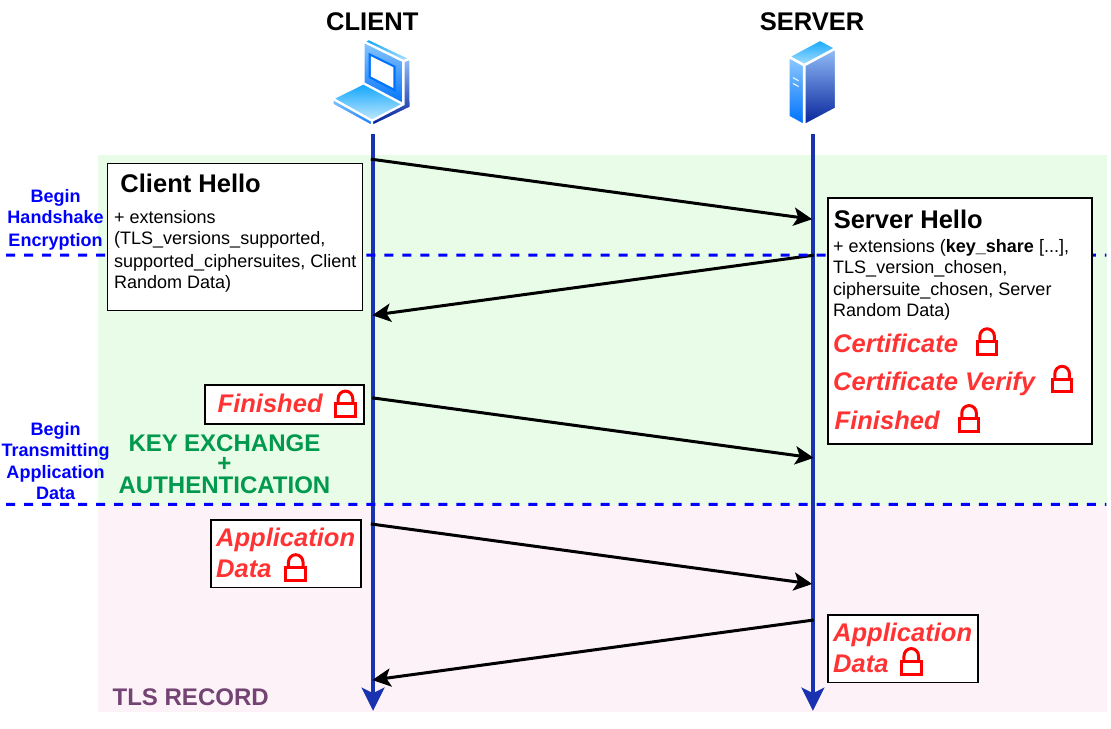}
    \caption{TLS 1.3 handshake protocol (plain messages in black, ciphered messages in red).}
    \label{fig:TLS1.3hs}
\end{figure}

In TLS 1.2, authentication and key exchange require two complete message round-trips -- as depicted in Figure \ref{fig:TLS1.2hs} --, while in TLS 1.3 require only one round-trip (since the Client's \textit{Finished} message can be sent among with the first \textit{Application Data} message), as depicted in Figure \ref{fig:TLS1.3hs}. This consolidation represents one of the main performance improvements.

As mentioned above, TLS performs authentication. More specifically, through the X.509 PKI \cite{RFC_5280}, which supports three authentication modes: server-only authentication, mutual authentication, and non-authentication (which is insecure and rarely used). The most widely adopted approach is server-only authentication. Although our scheme could be extended to support mutual authentication, we have opted to focus exclusively on the server-only authentication mode, as it is by far the most prevalent in practical TLS deployments.

The TLS 1.2 handshake begins with the \textit{Hello} messages, where the client and the server negotiate cryptographic parameters, including the communication ciphersuite, the version of the protocol, the public key algorithm, and two random values known as \textit{Client Random} and \textit{Server Random}, which are later used to derive the shared secret. Once these parameters are agreed on, the server initiates the authentication process by sending a \textit{Certificate} message, containing the certificate for this host (with the hostname, public key, and a signature from CA asserting that the owner of the certificate's hostname holds the private key associated to this certificate) or a list of further certificates, each of which signs the previous certificate, and which form a chain of trust leading from the host certificate to a trusted certificate that has been pre-installed on the client.

The server key exchange process begins with the generation of a pair of private and public keys for key exchange purposes. The server transmits its public key within the \textit{Server Key Exchange} message. Following this, the server sends the \textit{Server Hello Done}, indicating that it has completed its \textit{Hello} process.

When the client receives this message, it first performs the authentication verification and, only after the successful verification, begins its key exchange process. The client generates a pair of private and public keys and transmits the public key to the server inside the \textit{Client Key Exchange} message. With this information, the client calculates the encryption keys by applying the negotiated PKI algorithm to the server public key and the client private key, obtaining the \emph{pre-master secret}. The client then derives the \emph{master secret} from the \emph{pre-master secret} using an HMAC-based Pseudo-Random Function (PRF) among with the \textit{Client Random} and the \textit{Server Random}. From \emph{master secret}, all cryptographic keys are derived using the same HMAC-256 function: the client MAC key, the server MAC key, the client write key, the server write key, the client write Initialization Vector (IV) and the server write IV. Following this, the client transmits a \textit{Change Cipher Spec} message to notify the server that subsequent records will be encrypted with the just derived keys. Up to this point, all shared information remains unencrypted. 

Upon receiving this message, the server computes the same set of encryption keys following the same process and sends its own \textit{Change Cipher Spec} message. To guarantee the success and integrity of the handshake, both parties calculate a hash of all previous handshake messages, encrypt them with the session key, and exchange them within the \textit{Finished} message. If both parties successfully decrypt and verify the hash, they confirm that the handshake is successful.

The TLS 1.3 handshake process differs significantly. Due to the considerable reduction in ciphersuites in TLS 1.3, the client anticipates that the server will select one of the five available. Consequently, the client proactively computes the necessary key pairs for each possible PKI algorithm, and transmits all resulting public keys within \emph{Key Share} message extension, which is included into the \textit{Client Hello} message. This message also specifies the TLS version (marked TLS 1.3), the list of supported ciphersuites (since only \texttt{TLS\_AES\_128\_GCM\_SHA256} and \texttt{TLS\_CHACHA20\_POLY1305\_SHA256} are required to be supported; this field may vary), and the \textit{Client Random}. Upon receiving the \textit{Client Hello}, the server selects its preferred ciphersuite, generates its own pair of public and private keys, and transmits its public key back to the client via the \textit{Key Share} extension of the \textit{Server Hello} message. This message also confirms the selected TLS version (TLS 1.3) and the chosen ciphersuite. Both the client and the server then use their own private keys and the received public keys to compute a shared \textit{pre-master secret}. In contrast to TLS 1.2, where the \textit{master secret} is derived from the \textit{Client Random} and \textit{Server Random}, TLS 1.3 computes the \textit{master secret} using the \textit{pre-master secret} and a SHA-384 hash of the \textit{Client Hello} and \textit{Server Hello} messages. The resulting \textit{master secret} is used as input to the HMAC-based Key Derivation Function (HKDF) -- a key distinction from TLS 1.2, which used HMAC-SHA256. Additionally, the server sends the \textit{Certificate} message, but unlike TLS 1.2, this message is encrypted. Although the authentication process follows a similar approach to TLS 1.2, encryption of the \textit{Certificate} message provides an additional layer of security.

The client decrypts the server's \textit{Certificate} received using the just derived keys, and verifies its identity, following a process similar to TLS 1.2. Upon successful verification, the server is authenticated and both parties perform the same verification with the \textit{Finished} message to confirm the integrity and success of the handshake.

The TLS 1.3 handshake requires fewer communication round-trips compared to TLS 1.2, resulting in a faster establishment of secure connections with reduced latency \cite{TLS1.3vs1.2}. This streamlined process enables devices of real-time applications to interact with servers in a smoother way. The reduction in the number of supported ciphersuites also contributes to faster execution. Moreover, TLS 1.3 also implements Zero Time Resumption (0-RTT), a feature that allows a TLS client to transmit data during the initial round-trip, thereby reducing the time required to establish a secure connection.

\subsubsection{TLS Record}

At the lowest layer of TLS lies the TLS record layer, which is responsible for transmitting data blocks between the client and the server. Each block can hold up to $2^{14}-1$ bytes of data \cite{RFC_2246}. Subsequently, each record is compressed and encrypted according to the algorithms negotiated during the TLS handshake.

There is a limit to the amount of data that can be securely encrypted with a single key. For AES-GCM, up to $2^{24.5}$ full-size records (approximately 24 million) can be encrypted on a given connection while maintaining a safety margin of around $2^{-57}$ for Authenticated Encryption (AE) security \cite{Paterson_24}. This limitation requires periodic updates to cryptographic keys in long-term connections that require the transmission of large amounts of data. Unlike TLS 1.3, TLS 1.2 does not provide a built-in mechanism for rekeying. As a result, there is no standardized approach for initiating key updates during an active session. To achieve a similar result using TLS 1.2, a renegotiation of the TLS connection is required, which involves executing a new handshake within the existing TLS session to establish new cryptographic parameters, including encryption keys. In contrast, TLS 1.3 enables both peers to request session key updates using a \textit{Key Update} message, which can be sent by either peer at any time after the \textit{Finished} message.

This mechanism allows either party to initiate a key update for the entire connection. Following transmission of a \textit{Key Update} message, the sender must use the next generation of traffic keys to encrypt subsequent messages, while the receiver must update its receiving keys accordingly to the Traffic Key Derivation process described in RFC 8446. Both parties must encrypt the \textit{Key Update} messages with the old keys, while all subsequent messages with the new keys.

\subsection{Related Work}

Given the growing need to modify the security infrastructure against quantum threats, numerous studies have investigated the integration of quantum-resistant authentication and key exchange mechanisms into the TLS protocol (Refs. \cite{Tanizawa, Paquin, Sikeridis, Tasopoulos, Mink, Hubermann, Dowling, Huang, Shim, Giron, CRubio_2, CRubio, CRubio_3}).

Some of these works remain theoretical, offering integration specifications without experimental implementations (Refs. \cite{Tanizawa, Mink, Dowling, Huang, Shim, Giron}). Ref. \cite{Hubermann} implement the proposed methods, but do not provide an analysis or evaluation of the results. A separate group implements these methods, but does not deploy them in production environments (Refs. \cite{Paquin, Sikeridis, Tasopoulos, CRubio_2, CRubio, CRubio_3}). These works also differ in their focus on TLS versions, with some targeting TLS 1.2 (Refs. \cite{Tanizawa, Hubermann, CRubio}), and others focusing on TLS 1.3 (Refs. \cite{Paquin, Sikeridis, Tasopoulos, Dowling, Huang, Shim, Giron, CRubio_2, CRubio_3}).

Among the previous work, some opt not to implement any form of authentication, relying instead on classical methods (Refs. \cite{Mink, Hubermann, Shim, Giron}). Others choose to implement Pre-Shared Keys (PSKs) obtained through QKD (Refs. \cite{Tanizawa, Dowling}). Furthermore, there are works that rely on PQC for authentication (Refs. \cite{Paquin, Sikeridis, Tasopoulos, Huang, CRubio_2, CRubio, CRubio_3}). The PQC algorithms used in these works differ, and some of them are not among the later standardized in FIPS 203, 204 and 205.

Regarding key exchange, some works delegate this task to PQC (Refs. \cite{Paquin, Sikeridis, Tasopoulos, Dowling}), while others rely on QKD (Refs. \cite{Tanizawa, Mink, Hubermann}). Additionally, certain studies propose a hybrid PQC-QKD solution (Refs. \cite{Huang, Giron, CRubio}), while others suggest the use of QKD in the network backbone and PQC for end-to-end communications (Ref. \cite{Shim}). Notably, only Refs. \cite{CRubio_2, CRubio_3} propose a combined approach that incorporates classical cryptography, PQC, and QKD for key exchange. Among the papers that advocate QKD or hybrid key exchange, only two propose the use of standards to obtain keys (Refs. \cite{Shim, CRubio_2}), although one of them has not been experimentally integrated into TLS (Ref. \cite{Shim}). Furthermore, none of the works utilize SDN-enabled QKD related standards. The use of standards is essential for real-world implementations. Otherwise, it would be limited to a specific QKD implementation, an unacceptable handicap for a real, scalable network, and a no-go for telcos looking seriously at its usage in production facilities. 

Upon reviewing the state of the art, and to the best of the author's knowledge, none of the existing works implement a hybrid solution that combines classical, QKD, and PQC while also integrating QKD key retrieval and SDN in compliance with ETSI standards. Furthermore, no solution combines these aspects with the ability to perform key updates within the TLS 1.3 framework or has been extensively tested and deployed in a real, large, heterogeneous SDN network deployed in production facilities (i.e., the Madrid Quantum Communications Infrastructure, MadQCI \cite{MadQCI}). Additionally, none of the existing solutions propose the use of a dedicated ciphersuite within the TLS 1.3 framework, which not only is an elegant solution to promote crypto-agility as well as a pathway for future evolution, but also ensures backward compatibility -- a critical feature to facilitate a smooth transition towards quantum-safe infrastructures. Our approach enables the selection of the key exchange mechanism through ciphersuites, in line with the core philosophy of the TLS standard.

We would like to emphasize the importance of making TLS quantum-safe \cite{ETSI_GR_QS_004}, and the hybridization efforts presented in papers like this one, since there are currently efforts to standardize a hybrid TLS \cite{TLS_hybrid_design}, which could benefit from works like the one presented here.
\section{Quantum-Safe TLS}\label{sec:Methods_v2}

In this section, we will provide details on the implementation of the proposed Quantum-Safe TLS.

The proposed architecture employs two quantum-resistant solutions to mitigate the quantum threat. QKD is utilized to establish a secure symmetric key between communication endpoints, ensuring confidentiality even in the presence of quantum adversaries. PQC is integrated to secure authentication during the TLS handshake, and to establish a shared secret between two endpoints, leveraging PQC's resistance to quantum computing attacks. Specifically, this work adopts FIPS 203 (ML-KEM) and FIPS 204 (ML-DSA) as the PQC algorithms of choice. To maintain classical security as a fallback, in case both quantum-resistant methods are compromised or vulnerabilities are discovered, the system employs a hybrid approach that combines quantum-safe methods with classical cryptography techniques, specifically DHKE. By incorporating DHKE alongside QKD and PQC, resilience is enhanced against unforeseen vulnerabilities in any single cryptographic algorithms. Additionally, by encapsulating these algorithms within ciphersuites, the architecture facilitates usage, promotes crypto-agility, and supports a smooth transition towards quantum-safe infrastructures. This approach guarantees long-term confidentiality, in particular, because QKD is not based on computational security, thereby providing an additional layer of security against potential future advancements in both quantum and classical computing.


We will revisit the TLS 1.3 handshake and record phases, providing a detailed explanation of the modifications introduced in each to facilitate the implementation of quantum-safe TLS.

\subsection{TLS handshake}\label{sec:TLS_hs}

Authentication is one of the primary challenges posed by quantum threats. To address this, both PQC and PSKs have been proposed for initial authentication. PKI provides significant flexibility and versatility, but it introduces the challenge of relying on non-ITS authentication for ITS encryption. However, the primary issue with QKD lies in initial authentication in the first round, which can be addressed using PQC PKI. After the first authentication, subsequent QKD rounds are authenticated using keys from the previous rounds, which given the QKD characteristics is an ITS primitive. Since this process only requires a few seconds, and this limits the probability of breaking this first authentication, even with a quantum computer, the use of a classical PKI has been proposed for this step in QKD \cite{Renner}.

\begin{figure}[htbp]
    \centering
    \includegraphics[width=.97\linewidth]{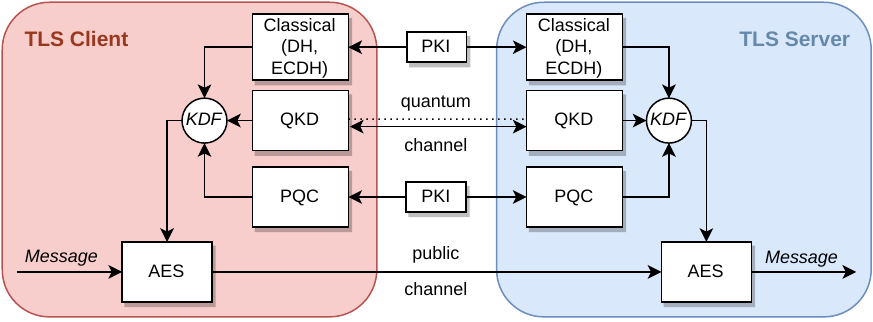}
    \caption{DHKE-QKD-PQC proposed hybridization. Shared secrets are generated for DHKE (whether conventional DH or ECDH), QKD, and PQC. A Key Derivation Function is applied to obtain the key to be used for ciphering. If high throughput is needed, the message can be ciphered using AES. If high security, potentially ITS, is the dominant requirement, an XOR can be used instead (see Ref. \cite{Renner}).}
    \label{fig:hybridization}
\end{figure}

Once authentication has been addressed, the keys obtained through different mechanisms can be combined using Key Derivation Functions (KDFs) that preserve the security properties. Figure \ref{fig:hybridization} illustrates the typical approach for classical cryptography, QKD and PQC hybridization \cite{QKD+PQC}, allowing the preservation of the ITS character of QKD if necessary. As QKD and PQC stand out as the two most prominent solutions to mitigate the quantum threat, significant efforts have been made to develop hybrid schemes \cite{QKD+PQC,QKD+PQC_2,Muckle+}.



Following this philosophy, we have introduced a new ciphersuite into the set supported by TLS 1.3 for any of the key exchange combinations involving DHKE, QKD, and PQC. Our focus will be on the ciphersuite \texttt{DHKE\_QKD\_PQC\_TLS\_AES\_256\_GCM\_SHA\_384}, which leverages the infrastructure of DHKE, QKD, and PQC to achieve authentication and key exchange, using AES-256 in Galois/Counter Mode (GCM) for encryption and SHA-384 for hashing.

All DHKE, QKD and PQC key exchanges require specific parameters, which are encapsulated within the \textit{Key Share} extension of the \textit{Client Hello} message, following a structure similar to that used for other supported ciphersuites. This allows the client to offer the necessary cryptographic material for each key exchange method, facilitating interoperability while preserving the security guarantees of the TLS 1.3 standard.

The QKD parameters are determined by the standards used for key extraction, with two primary standards currently available for this task: ETSI GS QKD 004 \cite{ETSI_004} and ETSI GS QKD 014 \cite{ETSI_014}. The first one operates at a lower level, offering reduced overhead. The latter is higher level and easier to use but requires more SW installed inside the QKD device and lacks important features like sessions or multi-domain addressing.

We could have used either of the two standard interfaces to retrieve the key from the northbound interface of the LKMS (refer to Figure \ref{fig:Node_KP} and Section \ref{sec:Arch} for further architecture details), and switching between them is straightforward in our implementation. However, we ultimately choose ETSI 004 due to its Quality of Service parameters and the concept of session establishment, which is aligned with the concept of TLS sessions. Note that from the southbound interface of the LKMS, for retrieving keys from the QKD modules, we can use either 004 or 014, depending on the standard that the manufacturer chose to expose the QKD keys.

\begin{figure}[htbp]
    \centering
    \includegraphics[width=.97\linewidth]{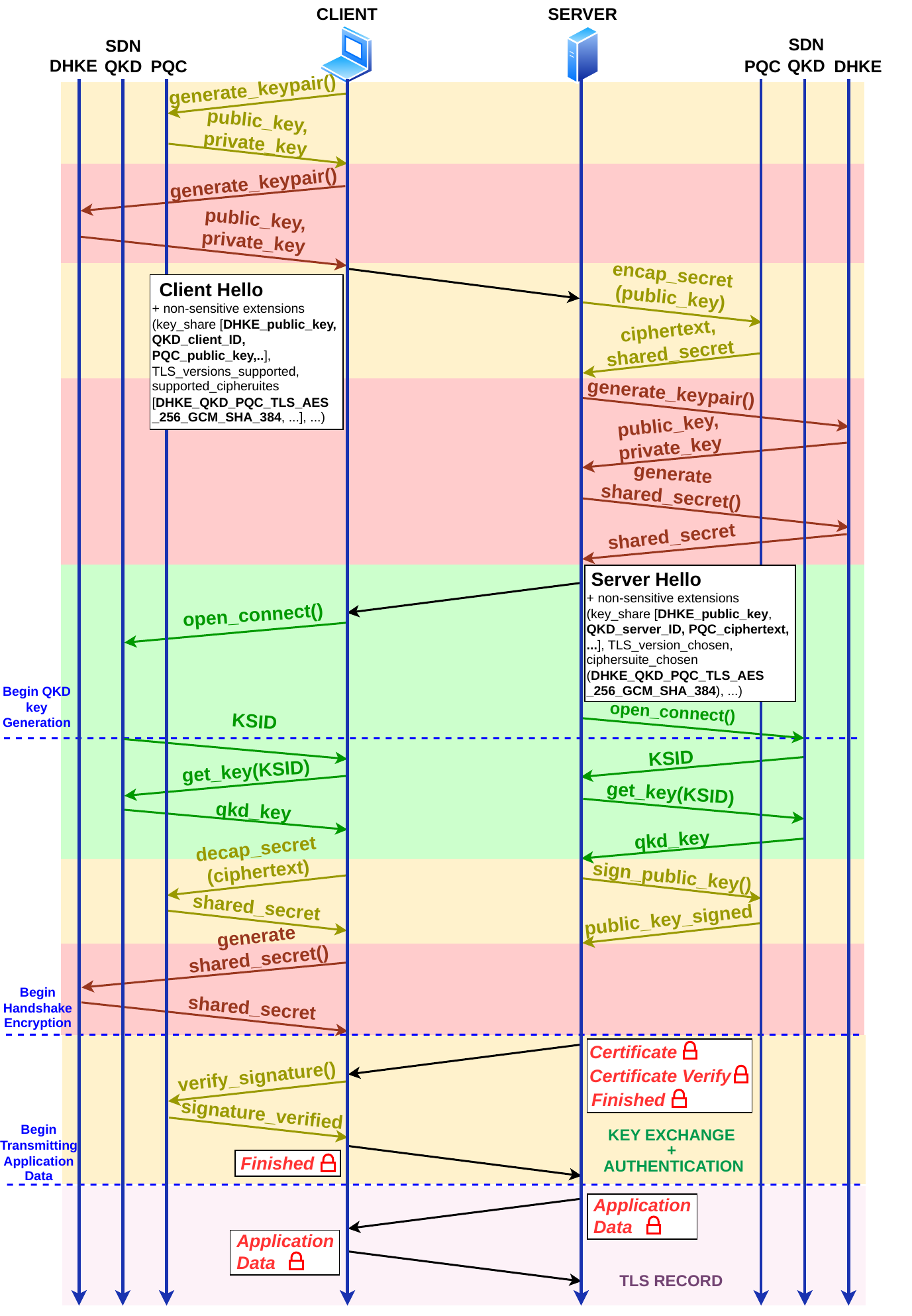}
    \caption{Quantum-Safe TLS 1.3 handshake protocol message exchange used in our implementation. Plain text messages are represented in black, while encrypted messages are in red. Message specific to ETSI GS QKD 004 are shown in green. PQC messages appear in mustard yellow, and DHKE messages in red.}
    \label{fig:PQC_QKD_TLS1.3hs}
\end{figure}

Figure \ref{fig:PQC_QKD_TLS1.3hs} illustrates the operation of the Quantum-Safe hybrid TLS handshake. Although the implementation could have been guided through the use of PSKs, our approach maintains consistency with the existing structure of TLS 1.3 by introducing a new ciphersuite within the supported set. This design enables a quantum-resistant solution with minimal modifications of the standard, promotes crypto-agility, and ensures backward compatibility.

Initially, the client generates private and public key pairs for both PQC and DHKE. The client then sends the \textit{Client Hello} message, which includes the \textit{Key Share} extension containing the client's DHKE public key, the QKD client application identifier, and the PQC public key. The preferred ciphersuite is configured as \texttt{DHKE\_QKD\_PQC\_TLS\_AES\_256\_GCM\_SHA\_384}. Upon receiving this message, the server computes the PQC shared secret and the corresponding PQC ciphertext. The server then generates its own DHKE key pair and computes the DHKE shared secret using the client's DHKE public key and the server's DHKE private key. Finally, the server constructs and transmits the \textit{Server Hello} message, embedding its own \textit{Key Share} extension, which contains the server's DHKE public key, the server's QKD application identifier, and the PQC ciphertext.

When the \textit{Server Hello} message arrives at the client, both endpoints initiate the standard ETSI 004 procedure by calling \texttt{open\_connect}, passing the QKD application identifiers as parameters. This request establishes a session and returns the Key Stream ID (KSID) associated with these identifiers, which is subsequently used by both parties to perform the \texttt{get\_key} request, retrieving the QKD key associated with the provided KSID. This process is optimized to significantly reduce the latency associated with the QKD request by overlapping both client and server QKD requests, ensuring a more efficient TLS handshake (as shown in Figure \ref{fig:PQC_QKD_TLS1.3hs}).

Once both parties have retrieved the TLS session key by using the KDF to hybridize the shared secrets of DHKE, QKD, and PQC, they retrieve the \textit{master secret} using the default TLS 1.3 procedure. Next, the authentication process begins. The server uses the ML-DSA algorithm to sign its ephemeral public key. This sign is then included into the \textit{Certificate} message, along with the server's hostname and the signature of the CA that issued the certificate, asserting that the holder of the certificate's associated hostname also possesses the corresponding private key. To ensure confidentiality, this message is encrypted using AES-256-GCM with the previously established session key. Afterwards, the server hashes all the handshake records exchanged up to that point, allowing the client to verify that the handshake has not been tampered with. Finally, the server sends both the \textit{Certificate} and \textit{Finished} messages, also encrypted.

At this stage, the client proceeds to authenticate the server's identity. First, it decrypts the \textit{Certificate} message using the session key. The client then verifies the server's certificate signature using ML-DSA and checks the server's hostname against the CA provided. Afterward, the client decrypts the \textit{Finished} message and verifies the integrity of the handshake. Next, the client computes a new hash of all previously exchanged messages, creates a \textit{Finished} message encrypted with the session key, and transmits it to the server. The server decrypts it and performs hash verification to validate the integrity of the handshake, resulting in the successful completion of the handshake. The verification of handshake integrity remains unchanged, benefiting from the security provided by the TLS standard.

\subsection{TLS record}

As explained previously, the TLS 1.3 protocol includes a record layer message known as \textit{Key Update}, which is used to inform the receiver that the sender is updating its cryptographic keys and requests the receiver to also update its keys.

To achieve a complete hybrid TLS, and without relying on the TLS 1.3 KDF, the behavior of both endpoints must be modified upon receiving the \textit{Key Update}. When one endpoint sends this message, it has to be ensured that both parties acquire a new, non-derived DHKE-QKD-PQC session key for encrypting subsequent traffic. To support this, the \textit{Key Update} message is expanded to include a new PQC public key (from the initiator to the receiver) and a new DHKE public key. Additionally, the receiver must transmit a new PQC ciphertext and DHKE public key back to the initiator. Since the QKD application identifiers remain unchanged, these parameters shall not be included in this message.

\begin{figure}[htbp]
    \centering
    \includegraphics[width=.97\linewidth]{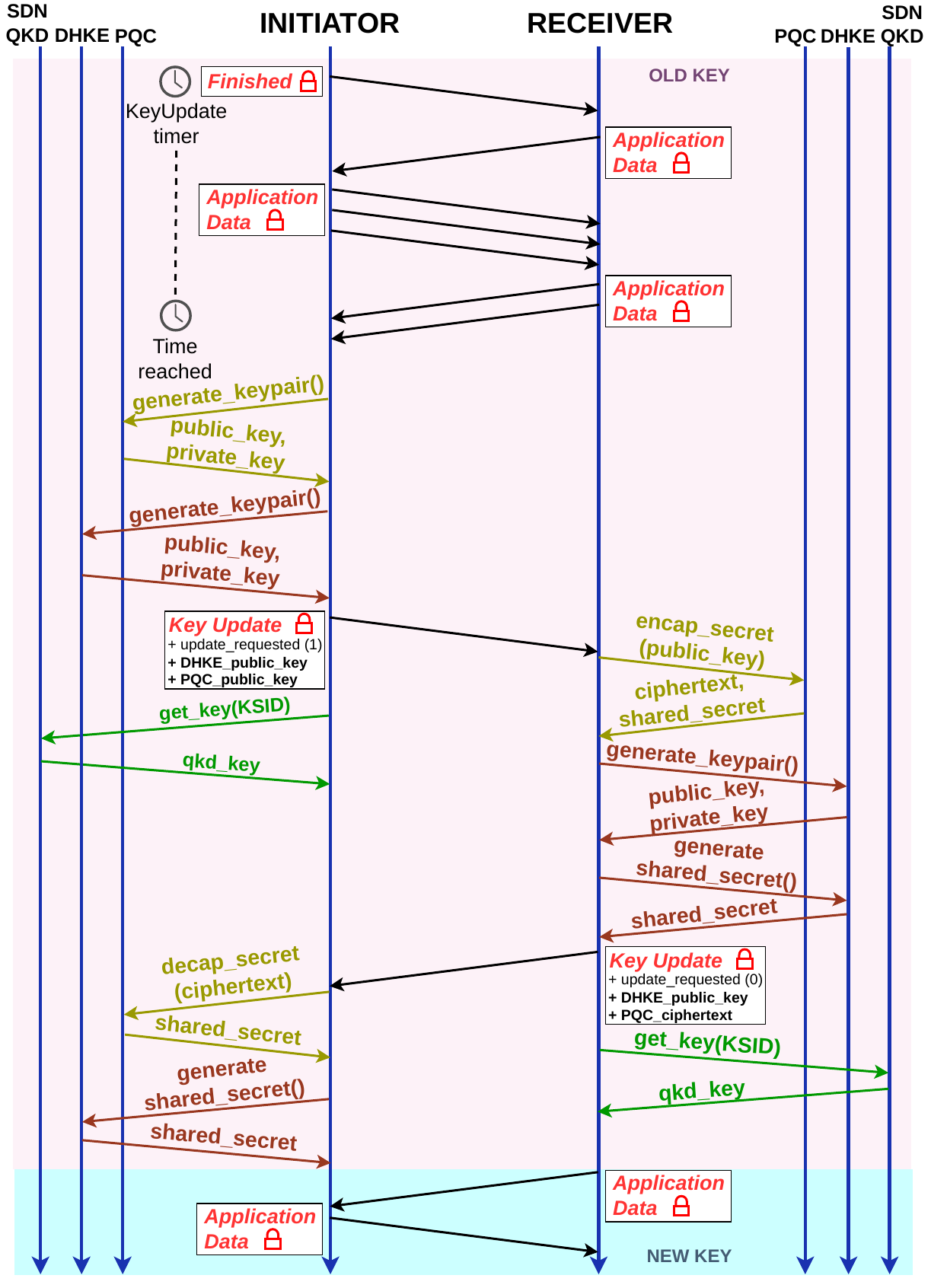}
    \caption{Key Update message operation. The pink background shows the exchange of messages encrypted with the old key. In blue, messages encrypted with the new key.}
    \label{fig:KP_PQC_QKD_TLS_KU}
\end{figure}

The key refresh process is illustrated in Figure \ref{fig:KP_PQC_QKD_TLS_KU}. Once encryption begins using the hybrid session key, a timer is triggered. Upon expiration, the initiator generates two new pairs of DHKE and PQC keys and sends a \textit{Key Update} message, embedding the PQC public key alongside the DHKE public key. This message is encrypted with the previous session key, following the specifications of RFC 8446, and the \texttt{request\_update} flag is set to 1, indicating the initiator's intention to refresh the session key. The initiator's QKD key request is executed concurrently with the transmission of the message, and DHKE and PQC key exchange operations are performed on the receiver's side.

Upon receiving the \textit{Key Update} message, the receiver encapsulates a secret using the provided PQC public key, generating a PQC shared secret and ciphertext, similar to the initial negotiation phase. The receiver also generates a new DHKE public key to be sent to the initiator and computes the corresponding DHKE shared secret. Both the PQC ciphertext and the new DHKE public key are transmitted back to the initiator within a new \textit{Key Update} message, with the \texttt{request\_update} flag set to 0, indicating that the receiver is not requesting to initiate another key refresh from the sender. The receiver then initiates a QKD key request executed concurrently with the message transmission, and the initiator's DHKE and PQC shared secret recovery. This approach minimizes the overall latency.
\section{Quantum-Safe SDN-TLS architecture}\label{sec:Arch}

In this work, our objective is to progress in the creation of a quantum-safe security infrastructure, focusing on how to make TLS quantum-safe, easily deployable, and independent of the underlying network model. However, to hybridize with QKD keys, these keys must be readily available. Given this requirement, it is logical to integrate the solution with a network model that supports QKD in the best possible way. As mentioned in Background, SDN is the \textit{de facto} choice in modern networks. Its flexibility greatly facilitates the integration of QKD in telecommunications networks, providing easy access to the QKD keys that must be used to create a hybridized key in quantum-safe TLS. In SDN, the logically centralized controller is the main piece that facilitates the configuration of the network resources to optimize its usage: managing network topologies, optimizing traffic flows, facilitating the implementation of new services and applications, etc. Consequently, our focus shifts to the connection between the SDN controller and TLS.

\subsection{TLS in a SDN: use of standards}

As outlined in the Quantum-Safe TLS section (\ref{sec:Methods_v2}), our approach for key extraction from QKD devices to LKMS relies on the main industry standards: ETSI GS QKD 004 \cite{ETSI_004} and ETSI GS QKD 014 \cite{ETSI_014}. The choice of standards depends on the interface provided by the QKD vendor. For key extraction from the LKMS to applications, we have preferred the use of ETSI 004 for various reasons. While both standards provide the same core functionality, their suitability is independent of the specific network model used, making ETSI 004 a practical choice for this purpose.

ETSI has also pursued the integration of QKD solutions into SDN paradigms. In March 2018, the work to establish the QKD Control Interface for Software-Defined Networks, known as ETSI GS QKD 015, started and was approved in March 2021 \cite{ETSI_015_1}, later upgraded in April 2022 \cite{ETSI_015}. This framework defines a structured approach for incorporating QKD into disaggregated network control plane architectures, particularly within SDNs, specifying abstraction models and workflows for interactions between SDN nodes (particularly the SDN agent) and the SDN controller.

\begin{figure}[htbp]
    \centering
    \includegraphics[width=.97\linewidth]{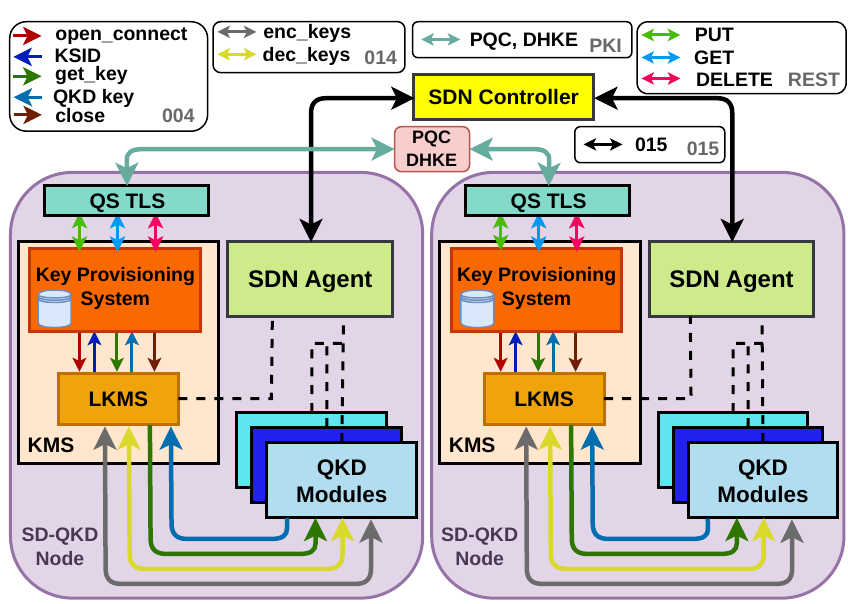}
    \caption{Operation of Quantum-Safe TLS in a SD-QKD network with two SD-QKD nodes. The Quantum-Safe TLS application communicates with the Key Provisioning System (KPS) using REST requests. The KPS, which is part of the KMS layer, communicates with the Local Key Management System (LKMS, following the ETSI nomenclature) via ETSI 004 requests. The LKMS communicates with the QKD Modules via ETSI 004 and ETSI 014 requests, depending on the interface the QKD vendor implements. The SDN agent communicates with the SDN controller using ETSI 015 requests, which are in charge of configuring the entire QKD-SDN network to communicate the two endpoints. Both Quantum-Safe TLS applications communicate through PQC \& DHKE PKI to retrieve a shared secret.}
    \label{fig:Node_KP}
\end{figure}

Figure \ref{fig:Node_KP} shows an abstract representation of an SD-QKD network with two SD-QKD nodes, illustrating their interaction during the proposed TLS operation. This Figure shows a layered model consisting of the Application Layer (Quantum-Safe TLS), the KMS Layer (Local Key Management System, LKMS \& Key Provisioning System, KPS), the Control Layer (SDN agent and controller) and the Quantum Layer, which may include one or more QKD modules, together with the associated quantum and service channels. The LKMS communicates with the Application Layer using ETSI 004, and with the QKD Modules using both ETSI 004 and 014. The SD-QKD node communicates with the SDN controller via the SDN agent using ETSI 015, enabling the retrieval of information from the QKD domain and supporting dynamic remote configuration of QKD systems.

This approach enables the creation, removal, and updating of QKD links between remote secure locations, allowing the establishment of end-to-end TLS connections among any two endpoints in the network. This process ensures that the QKD key is available to be hybridized with the other keys in the quantum-safe connection. To guarantee the timely provision of QKD keys and minimize the impact of latencies arising in the quantum plane, an efficient KPS is essential.

\subsection{Performance increase: QKD Key Provisioning}

Key provisioning refers to the secure distribution of cryptographic keys to authorized entities for use in secure communication. In the context of QKD, key provisioning is especially critical due to the variation in key rates, which depend on factors such as optical absorptions in the network and noise in the quantum channel, that can fluctuate significantly in a real-world dynamical network. Because the key material is a valuable and expensive resource, its usage must be optimized for efficiency \cite{KP}.

By implementing key provisioning into an SDN architecture, we leverage a key functionality that can significantly reduce deployment latencies and accelerate deployment: provisioning of QKD keys across the SDN network. By deploying the SDN network and using the ETSI 015 and 004 interfaces, each network node can provision a key stream, effectively minimizing the latency associated with TLS execution to near-zero levels (in our implementation, limited to executing a REST GET request to an internally executed service within the SDN node).

We have developed a REST service called the KPS, which operates internally within the KMS layer to provision QKD keys and facilitate key requests between the LKMS and applications. Upon network deployment, critical operations -- such as the ETSI 004 \texttt{open\_connect} -- are initiated, allowing all nodes in the network to generate key streams with any other network node. When an application, such as the Quantum-Safe TLS depicted in Figure \ref{fig:Node_KP}, requires a QKD key, it does not directly request it from the LKMS via the ETSI 004 interface. Instead, it sends a REST GET request to the KPS, specifying the neighboring node against which the key is required. The KPS then triggers the \texttt{get\_key} call from ETSI 004 against the LKMS. Since the LKMS has already pregenerated the key, this request incurs nearly zero latency, the only delay being the REST-API request latency. While ad-hoc implementations could further reduce latency at the cost of flexibility, we have prioritized a scalable approach over marginal performance gains (as can be seen in Table \ref{tab:total_004}, the maximum could be approximately 50ms).

\begin{figure}[htbp]
    \centering
    \includegraphics[width=.97\linewidth]{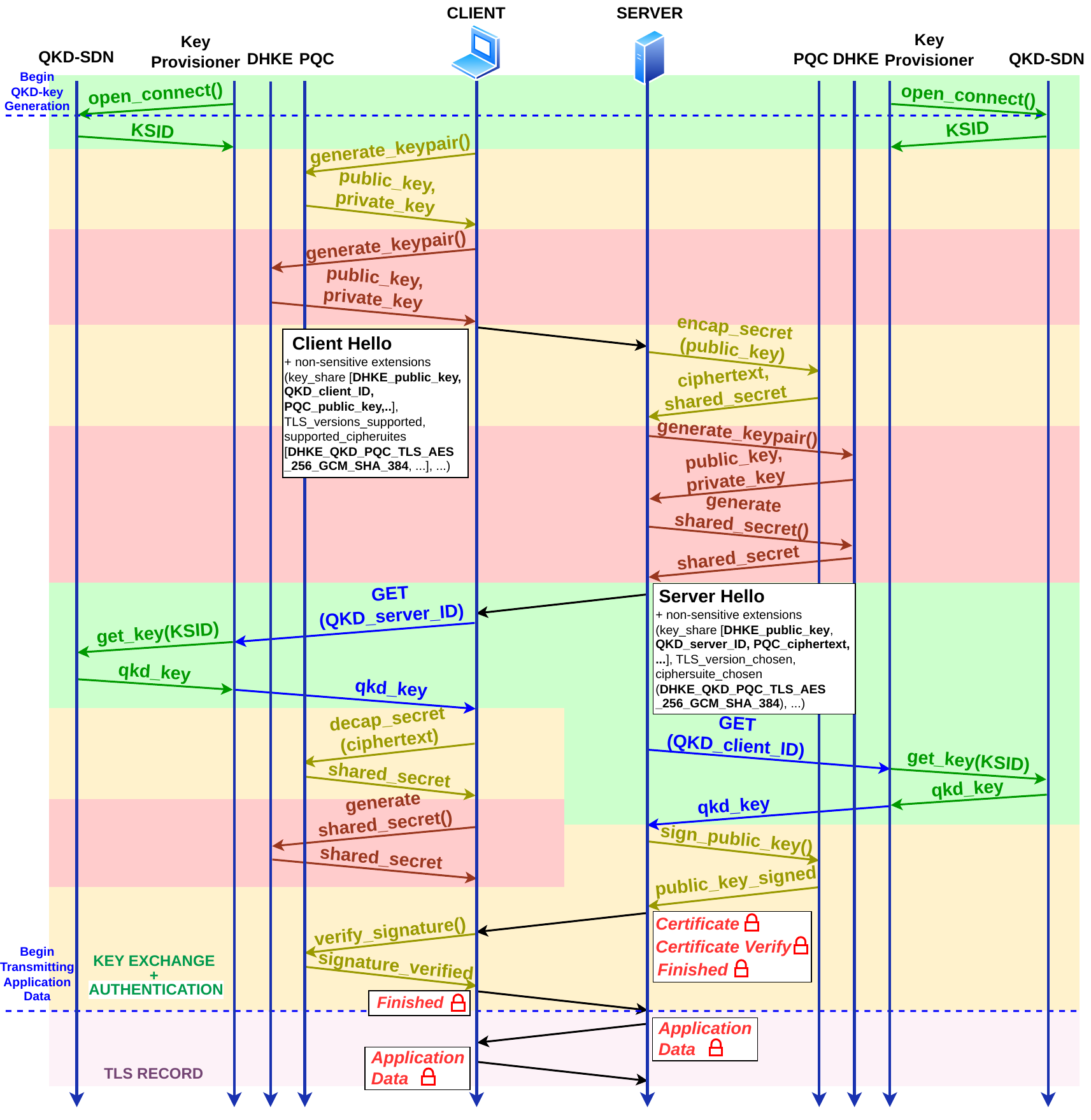}
    \caption{Quantum-Safe, SDN supported, Key Provisioned TLS 1.3 handshake protocol message exchange proposal. Plain text messages are represented in black, while encrypted messages are in red. Messages specific to ETSI 004 are shown in green, and REST requests are depicted in blue. PQC messages appear in mustard yellow, and DHKE messages in red. The key exchange and authentication phase proceeds up to the beginning of application data transmission, which is depicted in light pink.}
    \label{fig:KP_PQC_QKD_TLS1.3hs}
\end{figure}

Figure \ref{fig:KP_PQC_QKD_TLS1.3hs} shows the Quantum-Safe TLS handshake with SDN-supported QKD key provisioning. This version introduces a significant improvement compared to the previous implementation by allowing both client and server to send their \textit{Hello} messages after performing the ETSI 004 \texttt{open\_connect} request, decreasing the latency significantly, since most of the latency associated with ETSI 004 is attributed to this function -- around 99\% (inferred from Table \ref{tab:total_004}). This is due to the blocking nature of the \texttt{open\_connect} function in the ETSI 004 interface \cite{ETSI_004}, which requires the client to know the server's QKD application ID before making the request. In this approach, the SDN performs the function during network deployment. As will be shown in the Results section (\ref{sec:Results_v2}), key provisioning plays a fundamental role in the overall performance of quantum-safe TLS in networks.

\subsection{Key Transport in a SDN}

The main functionality of a QKD network is to securely transport keys between any nodes within the network. Therefore, we provide a brief introduction to this concept, as it forms a crucial foundation for the subsequent performance evaluation.

In an SDN environment, key transport refers to the secure transmission of cryptographic keys between different network elements or entities within the SDN infrastructure. Note that we are using the slightly broader concept of key transport, consistently with the use of TLS, albeit formally we are creating a quantum-safe key from key-agreement primitives. SDN enhances key transport by centralizing key management and distribution, leading to more efficient and scalable key provisioning. This centralization enables secure communication throughout the network infrastructure.

Due to physical limitations in the QKD key agreement process, it is often necessary for a ``logical'' link between two endpoints to span multiple physical links. For example, when the two endpoints are connected with an optical channel that has higher losses than those tolerable by the QKD devices, it is necessary to use intermediate nodes within reach. Thus, several jumps might be needed to cover the full distance. Given the distributed nature of this process and the involvement of intermediate devices, potentially from different manufacturers and using different QKD protocols and software components, key transport becomes a complex and non-trivial task in large QKD networks. The SDN controller must take all this into account and manage the possible links while ensuring that the complexity of the underlying network is transparent to network applications.

\begin{figure}[htbp]
    \centering
    \includegraphics[width=.97\linewidth]{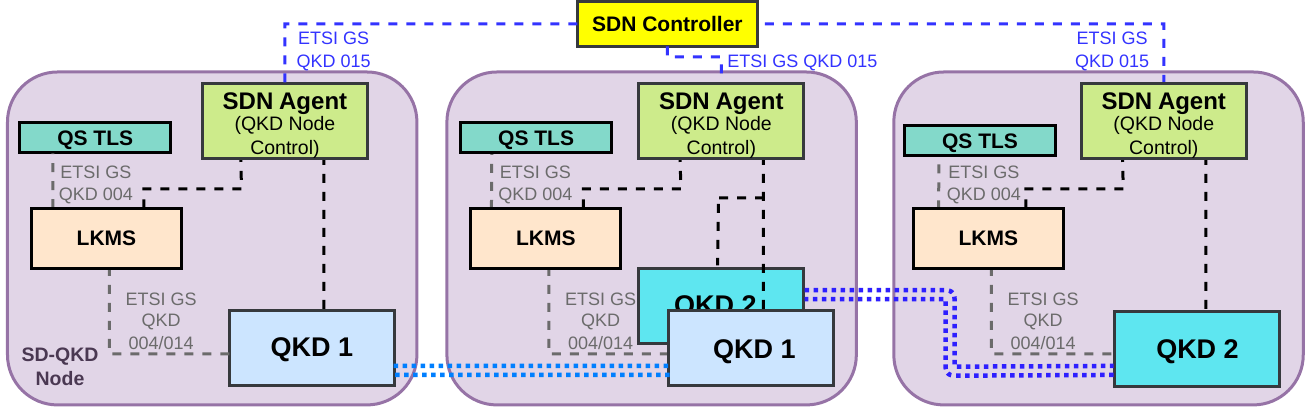}
    \caption{Example of Key Transport in a 3-node SD-QKD network scenario involving two QKD links.}
    \label{fig:KeyTransport}
\end{figure}

Figure \ref{fig:KeyTransport} illustrates the operation of a simple hop-by-hop key transport in a SD-QKD network. First, Nodes 1 and 2 establish a shared QKD key ($CK_0$), while Nodes 2 and 3 agree on a different QKD key ($CK_{1}$). Next, Node 1 generates a QRNG key ($K_{ENC}$), which it encrypts using the QKD key shared with Node 2 ($KN_{0} = K_{ENC} \oplus CK_{0}$). The encrypted key ($KN_0$) is transmitted to Node 2, where it is decrypted using the shared key ($K_{ENC} = (KN_0 \oplus CK_0)$). Node 2 then encrypts $K_{ENC}$ with the QKD key shared with Node 3 ($KN_1 = (K_{ENC} \oplus CK_1)$) and forwards the result to Node 3. Upon receiving the encrypted message, Node 3 decrypts it using its QKD key ($K_{ENC} = KN_1 \oplus CK_1$), thus retrieving the original QRNG key ($K_{ENC}$). Note that other schemes, more sophisticated than this simple encrypt/decrypt at each node mechanism, are possible. 

\subsection{MadQCI network}\label{sec:MadQCI}

The developments described above have been deployed in a quantum metropolitan communication testbed designed to be as close to a real-world, integrated quantum-classical network as possible. The MadQCI is an SDN heterogeneous network comprising 28 QKD modules from 4 different manufacturers \cite{MadQCI}. Deployed in production facilities, MadQCI supports the coexistence of classical and quantum signals within the same physical infrastructure. The network contains two domains, belonging to two infrastructure providers and connected by a border link. This setup supports the creation of QKD keys among any two nodes and has been running for almost three years in different configurations.

\begin{figure}[htbp]
    \centering
    \includegraphics[width=.97\linewidth]{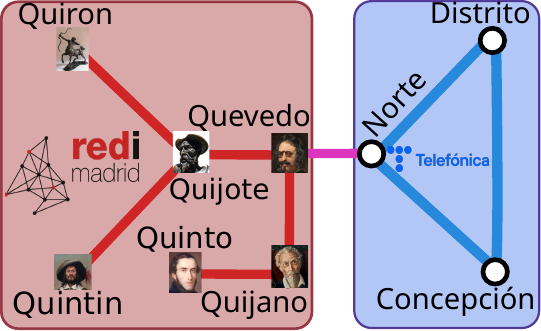}
    \caption{MadQCI network scenario, involving two network provider domains: Telef\'onica (depicted in blue) and RediMadrid (depicted in red). The border node is shown in purple.}
    \label{fig:MadQCI}
\end{figure}

Figure \ref{fig:MadQCI} presents a map of the MadQCI network. The MadQCI has been running a variety of applications with very different requirements. Some are highly sensitive to latency, while others have to support very large encrypted data throughput. This requires high key refresh rates and an efficient Key Provisioning System.

The proposed TLS implementation has been successfully developed and deployed in the MadQCI network for more than two years. It is noticeable that this network is a multi-domain network, and TLS has allowed to establish an end-to-end secure communication between any node in this network.

\section{Experimental results and discussion}\label{sec:Results_v2}

In this section, we present the performance evaluation using DHKE\footnote{For the sake of convenience, we will use DHKE as a general term to refer to any type of Diffie-Hellman Key Exchange. Specifically, DH will denote the variant based on discrete logarithms over prime fields, while ECDH will refer to the variant based on elliptic curves.}, QKD, PQC, and their various hybrid configurations, providing a comparative analysis of these security mechanisms, focusing on their latency, throughput, and overall impact on the proposed quantum-safe SDN-enabled TLS.

We emphasize the importance of a well-designed Key Provisioning System, as the underlying QKD network can introduce significant variability due to fluctuations in the quantum channel. These fluctuations depend on numerous factors, including the specific implementation of QKD, the routing of quantum signals, and how they share the physical infrastructure. While these factors do not directly influence the hybridization process, they can affect its overall performance. To ensure a stable environment that allows meaningful conclusions about the cost of hybridization, the Telef\'onica domain (shown in blue in Figure \ref{fig:MadQCI}) was chosen as the testing ground. This domain offers a controlled and limited setting, which makes it a better choice for obtaining isolated, reliable data. Since QKD systems continuously generate keys and the KPS buffers them, our study operates under the assumption of a properly dimensioned QKD network (i.e., the system avoids in situations where no QKD keys are available, eliminating delays caused by waiting for the QKD system to generate keys). Performance evaluation focuses on measuring the overhead introduced by hybridization, encompassing algorithmic and KPS-related factors, a realistic scenario for a correctly dimensioned QKD network.

Following an analysis of the state of the art in classical and quantum-safe key exchange technologies, summarized in Table \ref{tab:SecurityBits}, the following methods were selected for performance evaluation: DH 3072-bit (NIST group 15), ECDH curve P-256 (NIST group 19), ML-KEM-512, and QKD-128 bits. We included two commonly used DHKE methods for classical key exchange to assess their performance and evaluate their suitability for hybridization. The notable performance differences among these methods highlight their varying suitability for hybridization scenarios. The selected key exchange mechanisms were carefully chosen to align with equivalent NIST-defined security levels, ensuring a fair and consistent basis for a comparative performance analysis. This study adopts NIST Level 1 security, which corresponds to 128 security bits. This level is equivalent to the computational strength of a 128-bit exhaustive search, such as a brute-force attack on AES-128, representing a practical and widely accepted baseline for secure communications, balancing robust security and performance efficiency. Although the proposed model is prepared to accommodate higher security levels, the extensive statistical data gathered for Level 1 provides sufficient insights, and transitioning to Level 3 or Level 5 would significantly increase the time required for data collection.

For authentication, we used ML-DSA-65, which corresponds to NIST Level 3. Although alternatives such as ML-DSA-44 (NIST Level 2) or ML-DSA-87 (NIST Level 5) were available, ML-DSA-65 was chosen as an optimal balance between security and performance.
 
\renewcommand{\arraystretch}{1,3}
\begin{table}[htbp]
    \centering
    \caption{Asymmetric cryptography security analysis.}
    \begin{tabular}{ | c | c | c | c |}
        \hline Algorithm & Key sizes & NIST security & Security bits \\ \hline
        \multirow{5}{*}{DHKE \& RSA} & 1024 & Obsolete & 80 \\
        & 2048 & Avoid & 112 \\ & \textbf{3072} & \textbf{Level 1} & \textbf{128} \\ & 7680 & Level 3 & 192 \\ & 15360 & Level 5 & 256 \\ \hline
        \noalign{\hrule height 1pt}
        \multirow{6}{*}{ECDH} & 160 & Obsolete & 80 \\
        & 192 & Obsolete & 96 \\ & 224 & Avoid & 112 \\ & \textbf{256} & \textbf{Level 1} & \textbf{128} \\ & 384 & Level 3 & 192 \\ & 521 & Level 5 & 256 \\ \hline
        \noalign{\hrule height 1pt}
        \multirow{3}{*}{ML-KEM} & \textbf{512} & \textbf{Level 1} & \textbf{128} \\ & 768 & Level 3 & 192 \\ & 1024 & Level 5 & 256 \\ \hline
        \noalign{\hrule height 1pt}
         \multirow{3}{*}{QKD} & \textbf{128} & \textbf{Level 1} & \textbf{128} \\ & 192 & Level 3 & 192 \\ & 256 & Level 5 & 256 \\ \hline
    \end{tabular}
    \begin{tablenotes}
        \centering
        \item[1] DHKE, RSA \& ECDH data extracted from Ref. \cite{Paar_Understanding}, ML-KEM data extracted from FIPS 203.
        \item[2] As is standard in the field, the security-bits represent the equivalent strength of a cryptographic system against brute-force attacks.
        \item[3] The key exchanges selected for performance evaluation are highlighted in bold text.
    \end{tablenotes}
    \label{tab:SecurityBits}
\end{table}
\renewcommand{\arraystretch}{1}

We conducted two sets of experiments to evaluate the operation of our quantum-safe TLS solution with key transport: one using a direct connection and the other employing an intermediate QKD node. In the first set of experiments, the deployment was configured in the direct topology (Norte-Concepci\'on-Distrito nodes), with Norte acted as the TLS client and Concepci\'on as the TLS server. This setup established a direct connection between the TLS client and the TLS server, without key relay. In the second set, the deployment was configured in the key relay topology (Norte-Distrito-Concepci\'on nodes), where Norte acted as the TLS client, Concepci\'on as the TLS server, and the QKD key was transported through Distrito, acting as a trusted node. These two configurations were managed through the SDN controller.

To ensure the randomness of the experiment and prevent network conditions from influencing the performance results, the experiments were conducted by alternating each topology every 10 minutes over a two-week period. This scheduling aimed to capture diverse network states, providing robust and unbiased measurements. The same machines were used for the TLS client and server in both topologies to minimize external factors that could affect latency measurements. The security mechanism used in each test was randomly selected, alternating between DH, ECDH, QKD, PQC, and their combinations. In total, more than 1,500 instances were collected for each topology, with approximately 150 instances recorded for each combination of key exchange.

The QKD link in the direct topology spans 8.1 km of optical fiber with an attenuation of 2.9 dB and is secured using ID Quantique Cerberis 3 systems, which implement the Coherent One Way protocol with time-bin encoding \cite{MadQCI_20}. These systems comply with ETSI 014 for exposing QKD keys. In the key relay topology, the QKD network is composed of two QKD links. The first, connecting Norte and Distrito, spans 15 km of optical fiber with associated losses of 8.3 dB, while the second, connecting Distrito and Concepci\'on, spans 15.3 km with 14.3 dB of losses. Both links use Toshiba QKD devices that operate the T12 protocol \cite{MadQCI_21}, an optimized version of BB84. Similarly to Cerberis 3 systems, Toshiba devices comply with ETSI 014 for key export. Note that these links are exclusively dedicated to the quantum channel; the classical communication channel remains unchanged for both topologies.

\begin{figure*}[htbp]
    \centering
    \includegraphics[width=.97\linewidth]{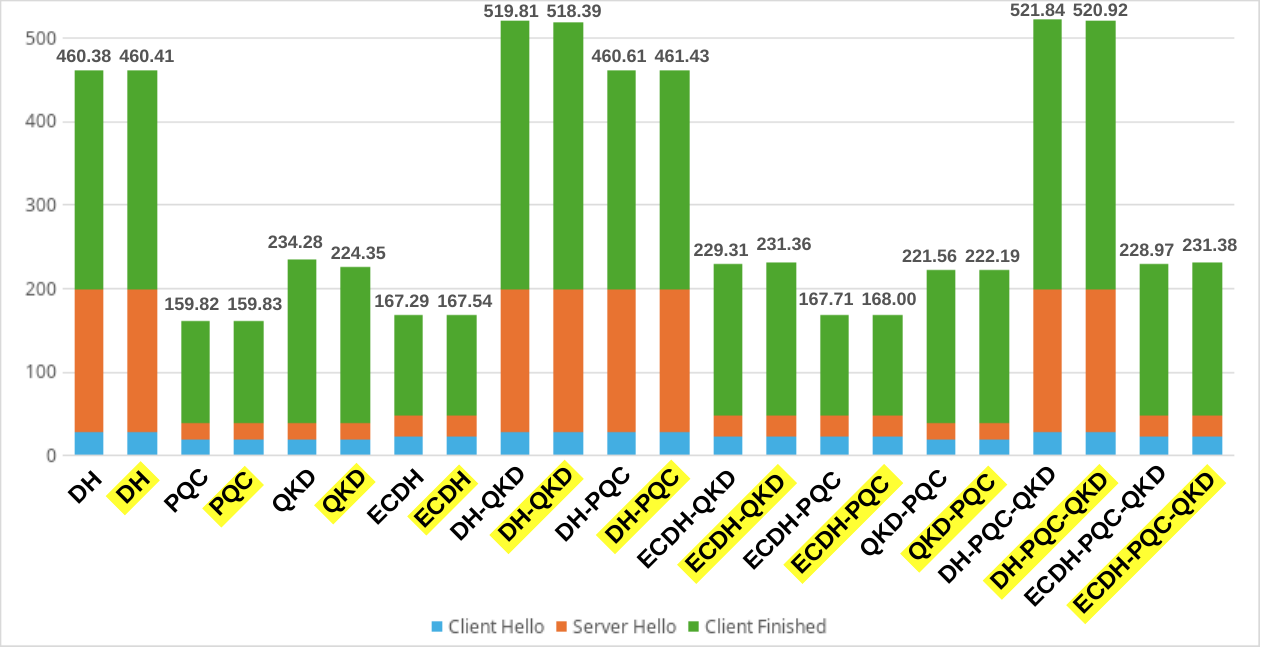}
    \caption{Time (in microseconds) taken by the TLS client to send the \textit{Client Hello} (in blue), to receive the \textit{Server Hello} (in orange), and for the TLS server to receive the \textit{Finished} message from the client (in green). The time is measured for each of the proposed key exchange combinations, in both direct and key relay topologies (with the former highlighted in yellow).}
    \label{fig:TLS_messages}
\end{figure*}

Figure \ref{fig:TLS_messages} shows the time associated with the TLS protocol for both topologies, measured from the initiation of the \textit{Client Hello} message sent by the client to the reception of the client's \textit{Finished} message by the server. Table \ref{tab:ncd_Results} shows the duration of each operation associated with the proposed key exchanges and authentication, along with an overall calculation of these times. The results shown in this table focus exclusively on the key relay topology, as the direct topology yields similar results due to the design of the KPS, which mitigates latency increases. Without this mitigation, the results would heavily depend on quantum network details, complicating the analysis of TLS performance in isolation. Table \ref{tab:total_004} presents the times corresponding to ETSI 004 requests on key retrieval from LKMS to the quantum-safe TLS application. 

Figure \ref{fig:TLS_messages} reveals that QKD performance remains unchanged in the key relay topology. The observed differences in TLS handshake times for key exchanges involving QKD (i.e., QKD, DH-QKD, DHKE-QKD, QKD-PQC, DH-QKD-PQC and DHKE-QKD-PQC) are minimal, with recorded variations of +9.93 ms (+4.24\%), +1.42 ms (+0.27\%), -2.05 ms (-0.89\%), -0.63 ms (-0.28\%), +0.92 ms (+0.18\%), and -2.41 ms (-1.05\%) compared to direct topology, respectively. These slight fluctuations may be attributed to inherent network conditions, as the provisioning of QKD keys by KPS minimizes the potential impact of QKD key transport on TLS performance.

From Table \ref{tab:ncd_Results}, several conclusions can be drawn. The latency associated with QKD is significantly higher than compared to ECDH, with QKD being 26.96\% slower. However, compared to DH, QKD shows a 50.19\% improvement in speed. The latency contribution of PQC is almost negligible, which demonstrates its computational efficiency and is consistent with the results of other studies \cite{Cano_24}. For the ECDH-QKD-PQC key exchange, 18.80\% of the total time is attributed to ECDH, 77.73\% to QKD, 1.63\% to PQC and 1.82\% to authentication. Although PQC's speed is promising, practical deployments must consider the increased latency due to the additional data that must be transmitted
(e.g., for ML-KEM-512, the public key size is 800 bytes and the ciphertext size 768 bytes), which is beyond the scope of this study. For QKD, the data overhead on the classical channels is negligible since the QKD keys are generated by the QKD network itself without imposing significant demands on the classical infrastructure. This analysis highlights the trade-offs between speed, security, and complexity among the evaluated methods, demonstrating that hybrid approaches incorporating QKD and PQC can be viable for quantum-safe communication.

From Figure \ref{fig:TLS_messages} and Table \ref{tab:ncd_Results}, it can be concluded that the additional latency required to achieve full protection against both quantum threats and currently known side-channel attacks, compared to a mature ECDH implementation, is only 62.86 ms. This increase represents a small fraction of the total handshake time (+37.48\%), which seems to be a reasonable trade-off considering the significant security enhancements achieved.

Table \ref{tab:total_004} reveals that the most significant variation is found in the \texttt{open\_connect} function, which is responsible for establishing connections between the Norte and Concepci\'on endpoints. However, the overall increase in latency remains negligible, amounting to 0.02 ms for the TLS client and 0.07 ms for the TLS server. While key generation is slower in the key relay topology compared to the direct topology, this penalty is effectively mitigated by using pre-provisioned QKD keys, a standard practice since QKD devices are continuously generating keys. However, it should be noted that under very intensive use, key stores may become depleted, which would negatively impact performance. The extent of this penalty would depend on a complex set of conditions, including the type of QKD device, the level of losses, and the noise introduced in the channel due to shared fibers and classical traffic. Modeling all of these possibilities is beyond the scope of this work.

\renewcommand{\arraystretch}{1,3}
\begin{table*}[htbp]
\centering
\caption{\bf TLS performance evaluation (in ms) in the key relay topology (Norte-Distrito-Concepci\'on).}
\label{tab:ncd_Results}
    \begin{center}
    \resizebox{\textwidth}{!}{
        \begin{tabular}{ | c | c | c | c | c | c | c | c | c | c | c | c | }
        \hline Operation & DH & PQC & QKD & ECDH & DH-QKD & DH-PQC & ECDH-QKD & ECDH-PQC & QKD-PQC & DH-QKD-PQC & ECDH-QKD-PQC \\ \hline
        ML-KEM KP & \gray & $0.28\pm0.12$ & \gray & \gray & \gray & $0.28\pm0.05$ & \gray & $0.28\pm0.16$ & $0.28\pm0.08$ & $0.28\pm0.23$ & $0.28\pm0.04$ \\ \hline
        DH KP & $8.55\pm0.92$ & \gray & \gray & \gray & $8.54 \pm 1.86$ & $8.55\pm0.98$ & \gray & \gray & \gray & $8.55 \pm 1.34$ & \gray \\ \hline
        ECDH KP & \gray & \gray & \gray & $3.61\pm0.16$ & \gray & \gray & $3.61\pm0.28$ & $3.59 \pm 1.19$ & \gray & \gray & $3.58\pm0.29$ \\ \hline
        QKD Request & \gray & \gray & $51.48\pm1.85$ & \gray & $51.77\pm1.98$ & \gray & $51.41\pm1.62$ & \gray & $51.26\pm1.88$ & $51.44\pm1.90$ & $51.97\pm1.62$ \\ \hline
        ML-KEM Decap & \gray & $0.46\pm0.05$ & \gray & \gray & \gray & $0.48\pm0.06$ & \gray & $0.48\pm0.05$ & $0.49\pm0.06$ & $0.50\pm0.05$ & $0.49\pm0.05$ \\ \hline
        DH SS & $168.76\pm2.46$ & \gray & \gray & \gray & $174.15\pm2.47$ & $167.56\pm2.40$ & \gray & \gray & \gray & $176.10\pm2.71$ & \gray \\ \hline
        ECDH SS & \gray & \gray & \gray & $4.76\pm0.46$ & \gray & \gray & $4.93\pm0.22$ & $4.59\pm0.65$ & \gray & \gray & $4.89\pm0.66$ \\ \hline
        ML-DSA Verify & $0.41\pm0.03$ & $0.37\pm0.12$ & $0.41\pm0.17$ & $0.40\pm0.02$ &$0.42\pm0.03$ & $0.38\pm0.02$ & $0.41\pm0.02$ & $0.36\pm0.11$ & $0.37\pm0.02$ & $0.38\pm0.27$ & $0.37\pm0.02$ \\ \hline
        \noalign{\hrule height 3pt}
        ML-KEM Encap & \gray & $0.31\pm0.21$ & \gray & \gray & \gray & $0.31\pm0.02$ & \gray & $0.31\pm0.27$ & $0.31\pm0.16$ & $0.31\pm0.02$ & $0.31\pm0.01$ \\ \hline
        DH KP \& SS & $150.58\pm1.51$ & \gray & \gray & \gray & $150.71\pm1.56$ & $150.68\pm1.94$ & \gray & \gray & \gray & $150.73\pm6.27$ & \gray \\ \hline
        ECDH KP \& SS & \gray & \gray & \gray & $4.11\pm0.99$ & \gray & \gray & $4.10\pm0.84$ & $4.11\pm2.90$ & \gray & \gray & $4.10\pm0.07$ \\ \hline
        QKD Request & \gray & \gray & $49.51\pm4.36$ & \gray & $49.26\pm52.76$ & \gray & $49.16\pm3.08$ & \gray & $49.15\pm0.16$ & $49.52\pm80.06$ & $49.36\pm2.78$ \\ \hline
        ML-DSA Sign & $0.90\pm0.11$ & $0.81\pm0.25$ & $0.85\pm0.37$ & $0.90\pm0.11$ &$0.89\pm0.13$ & $0.84\pm0.13$ & $0.87\pm0.11$ & $0.82\pm0.23$ & $0.85\pm0.11$ & $0.88\pm0.12$ & $0.85\pm0.21$ \\ \hline
        \noalign{\hrule height 5pt}
        DH & $327.88$ & \gray & \gray & \gray & $333.41$ & $326.79$ & \gray & \gray & \gray & $335.38$ & \gray \\ \hline
        ECDH & \gray & \gray & \gray & $12.48$ & \gray & \gray & $12.64$ & $12.28$ & \gray & \gray & $12.57$ \\ \hline
        ML-KEM & \gray & $1.06$ & \gray & \gray & \gray & $1.07$ & \gray & $1.08$ & $1.08$ & $1.09$ & $1.09$ \\ \hline
        QKD$^{(1)}$ & \gray & \gray & $51.48$ & \gray & $51.77$ & \gray & $51.41$ & \gray & $51.26$ & $51.44$ & $51.97$\\ \hline
        ML-DSA & $1.31$ & $1.18$ & $1.27$ & $1.31$ & $1.31$ & $1.22$ & $1.28$ & $1.18$ & $1.23$ & $1.27$ & $1.22$ \\ \hline
        Total & \textbf{329.20} & \textbf{2.24} & \textbf{52.75} & \textbf{13.79} & \textbf{386.48} & \textbf{329.08} & \textbf{65.33} & \textbf{14.54} & \textbf{53.57} & \textbf{389.18} & \textbf{66.86} \\  \hline
        \end{tabular}}
        \begin{tablenotes}
            \centering
            \item[1] The upper section (above the thick black line) represents the timing of the TLS process, with the TLS client depicted above the thin black line and the TLS server below it. The lower section illustrates the average time required to perform key exchanges and authentication for each key exchange.
            \item[2] DH refers to Diffie-Hellman (NIST group 15, 3072-bit prime), ECDH to Elliptic-curve Diffie-Hellman (NIST group 19, curve 256-bit), ML-KEM to ML-KEM-512, QKD to QKD-128 bits, and ML-DSA to ML-DSA-65.
            \item[3] KP refers to key pair generation, SS to shared secret generation, Encap to KEM encapsulation and Decap to KEM decapsulation.
            \item[4] $^{(1)}$ Since QKD requests are executed concurrently, the total time is determined by the latency associated to the maximum of both requests.
        \end{tablenotes}
    \end{center}
\end{table*}
\renewcommand{\arraystretch}{1}

\renewcommand{\arraystretch}{1,3}
\begin{table}[htbp]
\centering
\caption{\bf ETSI GS QKD 004 times (in ms) in both, direct and key-relay, topologies.}
\label{tab:total_004}
    \begin{center}
        \begin{tabular}{ | c | c | c | c | }
        \hline Topology & Role & Open Connect & Get Key \\ \hline
        \multirow{2}{*}{Direct} & Client & $65,73\pm11,94$ & $0,62\pm0,15$ \\
        & Server & $47,13\pm5,84$ & $0,61\pm4,00$ \\ \hline
        \multirow{2}{*}{Key Relay} & Client & $65,71\pm12,09$ & $0,62\pm0,18$ \\
        & Server & $47,20\pm7,22$ & $0,62\pm1,83$ \\ \hline
        \end{tabular}
    \end{center}
\end{table}
\renewcommand{\arraystretch}{1}

The results on key refresh performance can be directly inferred by combining the times required for each of the KEMs (Table \ref{tab:ncd_Results}), with the latency associated with the transmission and reception of the \textit{Key Update} message over the network.
\section{Conclusions and future work}\label{sec:Conclusions}

In this work, we propose the first quantum-safe hybrid implementation of TLS 1.3 using approved QKD standards for key extraction, as well as for SDN control, and supporting advanced features like rekeying. The implementation is conceived as a real solution towards fully quantum-safe infrastructures, hybridizing keys from conventional Diffie-Hellman, Post-Quantum Cryptography and Quantum Key Distribution protocols by using ciphersuites, which facilitates crypto-agility while ensuring compatibility with previous versions and facilitating a smooth transition towards quantum-safe infrastructures. Additionally, the implementation supports a QKD network-aware architecture with a scalable Key Provisioning System, enhancing performance in complex QKD networks with trusted nodes. This system has been developed and running for more than two years in a complex QKD network deployed in production infrastructures designed to mimic as closely as possible a real-world quantum-classical integrated network.

The results demonstrated that, using a correctly dimensioned QKD network, the performance penalty incurred by hybridizing the three methods is minimal. This finding underscores the viability of hybridization as a practical approach to transitioning current security infrastructures toward quantum-safe ones, protecting data transfers even in the long term.

In this work, we present a direct hybridization approach that combines keys from different sources into a unified keying material. Alternative approaches include nesting protocols, where each protocol utilizes a separate key. While this nested method may introduce performance disadvantages, it provides an additional layer of security by relying on different implementations. This diversity mitigates the risk of a single point of failure, which could potentially compromise the entire system in the future.

Future work includes aligning our approach with ongoing developments in hybrid TLS standardization efforts within IEEE, thereby promoting the integration of QKD and PQC into widely adopted security protocols and enhancing its practical usability in diverse network environments. 


The authors would like to thank projects MADQuantum-CM, funded by Comunidad de Madrid (Programa de Acciones Complementarias) and by the Recovery, Transformation and Resilience Plan-Funded by the European Union-(NextGenerationEU, PRTR-C17.I1); the EU Horizon Europe project "Quantum Security Networks Partnership" (QSNP), grant 101114043; and EuroQCI-Spain, DEP grant 101091638; the QUBIP, funded by the European Union under the Horizon Europe framework program under grant agreement number 101119746, and the PQ-REACT European Union’s Horizon Europe research and innovation programme under grant agreement number 101119547.

\bibliographystyle{IEEEtran}
\bibliography{sections/bibliography}

\end{document}